\documentclass[12pt]{iopart}
\expandafter\let\csname equation*\endcsname\relax
\expandafter\let\csname endequation*\endcsname\relax
\usepackage[cmex10]{amsmath}
\usepackage[font=footnotesize]{subfig}
\usepackage{amssymb}
\usepackage[dvips]{graphicx}
\usepackage{fixltx2e}
\usepackage[abbr]{harvard}
\usepackage[export]{adjustbox}
\usepackage{textcomp}

\begin{document}

\title{RegQCNET: Deep Quality Control for Image-to-template Brain MRI Affine Registration}
\author{Baudouin DENIS de SENNEVILLE$^{1}$, Jos\'e V. MANJ\'ON$^{2}$, Pierrick COUP\'E$^{3}$}
\address{$^1$ CNRS, University of Bordeaux, ``Institut de Math\'ematiques de Bordeaux'' (IMB), UMR5251, F-33400 Talence, France}
\address{$^2$ ITACA, Universitat Polit\`ecnica de Val\`encia, Camino de Vera, s/n, 46022, Valencia, Spain}
\address{$^3$ CNRS, University of Bordeaux, Bordeaux INP, ``Laboratoire Bordelais de la Recherche Informatique'' (LaBRI), UMR5800, F-33400 Talence, France}

\eads{\mailto{{bdenisde@math.u-bordeaux.fr}, \mailto{jmanjon@fis.upv.es}, \mailto{pierrick.coupe@u-bordeaux.fr}}}

\vspace{10pt}
\begin{indented}
\item[]July 2020
\end{indented}

\begin{abstract}Affine registration of one or several brain image(s) onto a common reference space is a necessary prerequisite for many image processing tasks, such as brain segmentation or functional analysis. Manual assessment of registration quality is a tedious and time-consuming task, especially in studies comprising a large amount of data. An automated and reliable quality control (QC) becomes mandatory. Moreover, the computation time of the QC must be also compatible with the processing of massive datasets. Therefore, an automated deep neural network approaches appear as a method of choice to automatically assess registration quality. 

In the current study, a compact 3D convolutional neural network (CNN), referred to as RegQCNET, is introduced to quantitatively predict the amplitude of an affine registration mismatch between a registered image and a reference template. This quantitative estimation of registration error is expressed using metric unit system. Therefore, a meaningful task-specific threshold can be manually or automatically defined in order to distinguish usable and non-usable images. 

The robustness of the proposed RegQCNET is first analyzed on lifespan brain images undergoing various simulated spatial transformations and intensity variations between training and testing. Secondly, the potential of RegQCNET to classify images as usable or non-usable is evaluated using both manual and automatic thresholds. During our experiments, automatic thresholds are estimated using several computer-assisted classification models (logistic regression, support vector machine, na\"ive bayes and random forest) through cross-validation. To this end we used expert's visual quality control estimated on a lifespan cohort of 3953 brains. Finally, the RegQCNET accuracy is compared to usual image features such as image correlation coefficient and mutual information.

Results show that the proposed deep learning QC is robust, fast and accurate to estimate affine registration error in processing pipeline.
\end{abstract}

\vspace{2pc}
\noindent{\it Keywords}: Quality Control, Image-to-template registration, Deep Neural Network.\newline

\maketitle

\section{Introduction}

A wide variety of processing pipelines have been proposed in the literature to make automatic brain image analysis possible. Spatial and intensity normalizations are usually necessary prerequisites for functional \cite{Cook2006} \cite{Song2011} or structural studies \cite{Wei2002} \cite{BrainPipeline}. These are commonly achieved using suitable algorithms designed for image-to-template registration \cite{Ants_Eval} \cite{JENKINSON2012} \cite{Collins1994}, inhomogeneity correction \cite{preprocessing3} \cite{Sled1998}, or intensity normalization \cite{MRI_normalisation} \cite{Friston1995}. 
A visual human inspection of the data after each step of the processing pipeline is commonly employed to detect possible problems in the outputs. This visual quality control (QC) is unfortunately not feasible when a huge amount of imaging data is involved (typically more than several thousands scans). Consequently, with the rise of large-scale datasets, recent efforts are dedicated to the development of reliable QC methods to detect pipeline failures \cite{QC} \cite{QC_neuro}. Although out of the scope of the current study, an increasing interest in registration QC is noticed in radiation therapy \cite{RegQC_RT_1} \cite{RegQC_RT_2}.

\begin{figure}[h!]
\begin{minipage}[b]{\linewidth}
\centering
\centerline{\includegraphics[width=\linewidth]{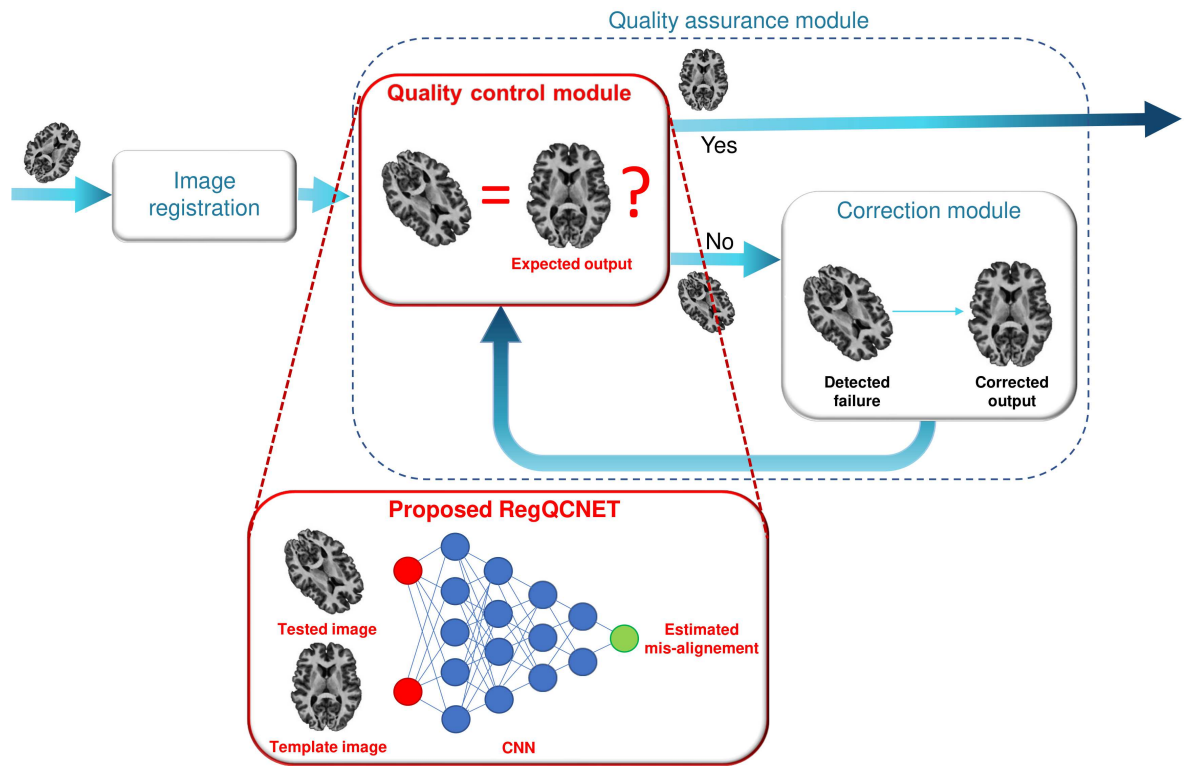}}
\end{minipage}
\caption{General principle of the proposed Quality Control (QC) on image-to-template registration. The current study aims at providing a QC module (red bock) designed to detect misaligned images. Note that this module can potentially be fed into an additional correction module (outside the scope of the current study).}
\label{fig:QC_scheme}
\end{figure}

Fig. \ref{fig:QC_scheme} summarizes the context of the current paper. Our study focuses on the image registration step, this step is a necessary prerequisite to co-register one or several brain scans onto a common space defined by a template image. In practice, mis-registered data are inherently encountered and thus a registration QC is needed (red box in Fig. \ref{fig:QC_scheme}) \cite{preprocessing4}. A manual assessment is generally employed for QC and thus automatic methods have been developed to achieve this task.  However, such manual strategy is time consuming. Random forest \cite{QC_RF} and convolutional neural network (CNN) \cite{LungQC_CNN} have been proposed to quantify registration accuracy for both parametric (\emph{i.e.}, rigid, affine) and deformable registrations in chest CT scans. In the context of neuroimgaging, several methods have been proposed for MRI brain registration to template space. In \cite{Fonov303487} a cross-entropy loss function is used as an objective function to train a deep neural network on a serie of 2D control images. This method produces qualitative estimation (\emph{i.e.}, good or not good) of rigid registration accuracy. In \cite{Bannister2019DeepNN} and \cite{Reg_DL}, a DICE metric between transformed and original organ contours is proposed as a surrogate of registration quality. The use of an indirect metric (\emph{i.e.}, DICE) estimated on an auxiliary task (\emph{i.e.}, segmentation) does not provide direct quantitative information on registration accuracy. Moreover, this information can be corrupted by segmentation error that is a complex task by itself. Finally, these three methods produce metrics (binary decision or auxiliary DICE) that cannot express registration error in metric unit system. Consequently, no meaningful task-specific threshold on the misalignment amplitude can be defined by a user.

Our contribution is four-fold:

\begin{enumerate}
 
 \item A compact 3D CNN is introduced to quantitatively estimate the quality of an affine alignment between a brain MRI and a template. The proposed QC network, referred to as RegQCNET, is quantitative and can be expressed using metric system units. Moreover, an efficient and robust training procedure, based on simulated affine transformations, is proposed. The inputs of the designed CNN are: the registered image and the reference template. Besides, it is demonstrated that RegQCNET meets computational requirements related to massive processing.

 \item The robustness of the proposed RegQCNET is analyzed on lifespan brain images undergoing various simulated spatial transformations and intensity variations between training and testing.
 
 \item The potential of RegQCNET to classify images as usable or non-usable is evaluated using both manual and automatic thresholds. Automatic thresholding is evaluated using several computer-assisted classification models (logistic regression, support vector machine, na\"ive bayes and random forest) through a cross-validation procedure. To this end we used expert's visual quality control estimated on a lifespan cohort of 3953 brains as a gold standard.
 
 \item The RegQCNET accuracy is compared to usual image features such as image correlation coefficient and mutual information.

\end{enumerate}

\section{Materials and Methods}

\subsection{Datasets}
\label{sssec:dataset}

\begin{figure}[h!]
\begin{minipage}[b]{\linewidth}
\centering
\centerline{\includegraphics[width=\linewidth]{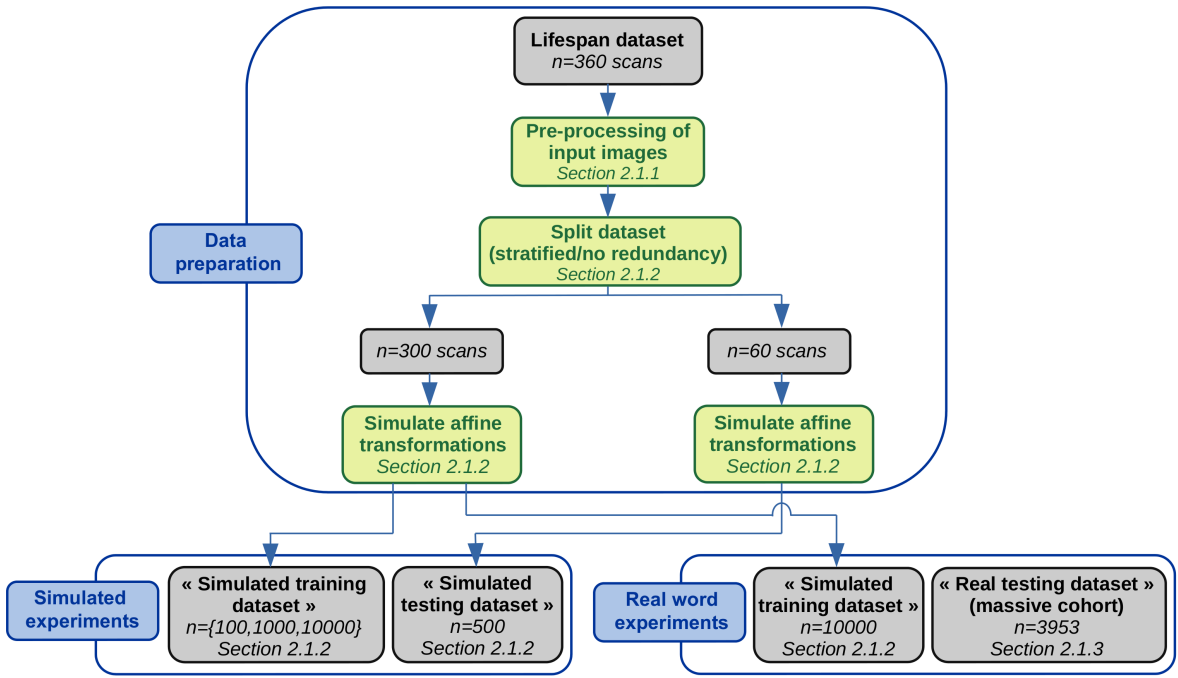}}
\end{minipage}
\caption{Processing sequence designed to generate datasets involved in experiments. 
Each generated dataset is displayed as a gray block. Image processing tasks are reported with green blocks.}
\label{fig:Scheme}
\end{figure}

Figure \ref{fig:Scheme} details the processing sequence designed to generate the datasets involved in our experiments. Throughout this study, we used 3 datasets — 1 for training and 2 for testing — referred to as ``Simulated training dataset'', ``Simulated testing dataset'' (these last two were built using a lifespan dataset, for which synthetic affine transformations were applied) and ``Real testing dataset''. 
First, all the native images have been downloaded from public databases. While considered as the native images, these images may contain specific preprocessing (i.e. NDAR includes defacing). Afterwards, all these images went through our preprocessing pipeline as described in the following.

\subsubsection{Preprocessing of input images.}
\label{sssec:preprocessing}

To ensure spatial and intensity normalization between images, we used a preprocessing pipeline. Consequently, all the scans involved in this study were first preprocessed beforehand using the volBrain pipeline \cite{preprocessing1}. This pipeline is based on the following steps: i) denoising \cite{preprocessing2}, ii) inhomogeneity correction \cite{preprocessing3}, iii) affine registration into the template space (181$\times$217$\times$181 voxels at 1$\times$1$\times$1 mm$^3$, the ICBM 152 Atlas template was taken as reference for registration \cite{ICBM152}), iv) manual human assessment of the registration as described in \cite{lifespan_dataset2} \cite{lifespan_dataset1}, and v) tissue-based intensity normalization \cite{preprocessing5}. Finally, image intensities were normalized using z-scoring using the mean and standard deviation from the complete image field-of-view.

\subsubsection{Simulated training and testing datasets.}

%\paragraph{The Lifespan dataset.}

360 T1-weighted MRI of cognitively normal subjects were randomly selected under constraints from the dataset used in our previous BigData study on normal aging \cite{lifespan_dataset2}. This dataset was based on 9 datasets publicly available (C-MIND, NDAR, ABIDE, ICBM, IXI, OASIS, AIBL, ADNI1 and ADNI2). From 1 to 90 years we selected 2 females (F) and 2 males (M) for each age (\emph{i.e.}, 2F and 2M of 1 year old, 2F and 2M of 2 years old and so on). Therefore, we obtained a balanced group with 50$\%$ of each gender uniformly distributed from 1 to 90 years. This balanced selection is done to limit bias introduction in training and testing datasets and to make our QC method robust to different age and gender.

All the 360 MRIs underwent a human quality control. Consequently, all these images were considered as correctly aligned with negligible residual registration mismatch. The RMSE was thus considered to be equal to 0. We are aware that these images are not perfectly aligned and that negligible errors might remain. Considering these remaining errors equal to epsilon or zero do not impact the rest of the study.

At this point, we have a set of 360 scans including spatial and intensity normalizations. This set was split into two separate datasets (stratified/no redundant data): 300 scans were used to build the ``Simulated training dataset'' and 60 scans were used to build the ``Simulated testing dataset''. This split was done under contraint to ensure well-balance of age and gender between both.

\paragraph{The ``Simulated training dataset''.}
\label{sssec:training}

One can expect that the training set has to be populated densely enough in terms of both anatomical inter-subject variability and simulated spatial deformations. During our experiments, RegQCNET was trained using $N$ simulated scans (randomly selected with replacement from the 300 above-mentioned scans). These scans were simulated using affine spatial transformations. In order not to favor large or small transformations, we force the RMSE distribution resulting from the simulated transformation to be uniform. We tested $N$-values in the following set: $\{100,1000,10000\}$. Note that the obtained training set may include several transformations for each patient when $N>300$.

\paragraph{The ``Simulated testing dataset''.}
\label{sssec:test1}

This set was composed of 500 scans (randomly selected with replacement from the 60 subjects selected to build the testing data). In this way, several transformations were applied to each patient ($\approx 8$ in average).

\paragraph{Simulated spatial affine transformations.}
\label{sssec:trans}

To train and test the proposed method, 3D spatial transformations --- composed by translation, rotation and scale variations --- were simulated. Ranges for X-, Y-, Z-translations, rotations around X-, Y-, Z-axis, X-, Y-, Z-scaling factors are detailed in the experimental section below. The RMSE was calculated for each simulated transformation. Let \verb"MAX_RMSE" be the upper limit of the simulated RMSE. A set of spatial transformations with a uniform RMSE distribution in the interval $\left[0,\right.$\verb"MAX_RMSE"$\left.\right]$ voxels was built. To this end, 3D spatial transformations composed by translation, rotation and scale variation, were simulated as follows:

\begin{itemize}
 
 \item X-, Y-, Z-translations were randomly selected separately in the interval $\left[-100,100\right]$ voxels,
 
 \item Rotations around X-, Y-, Z-axis were randomly selected in the interval $\left[-45,45\right]$ degrees,
 
 \item X-, Y-, Z-scaling factors were randomly selected in the interval $\left[0.5,1.5\right]$ (a factor of 1 being equivalent to no scaling),
 
\end{itemize}

These transformations were then applied using b-spline interpolation to the images (see Fig. \ref{fig:scans}).

\subsubsection{The ``Real testing dataset''.}

The performance of the proposed RegQCNET was evaluated on a massive database ($N=3953$) including cognitively normal patients, patients with Alzheimer’s Disease (AD) and Mild Cognitive Impairment (MCI). These 3953 MRIs were the remaining subjects from the large-scale cohort used in \cite{lifespan_dataset1} after removal of the 360 cognitively normal subjects used to build the simulated training and testing dataset. Consequently, the real testing dataset contained pathological alterations unseen in the training dataset. A visual assessment was done by checking screen shots of one sagittal, one coronal and one axial slice in middle of the 3D volume using the volBrain reports \cite{preprocessing1}. Therefore, a human-based QC was available for all the scans and was used as qualitative ground truth.

\subsection{Proposed RegQCNET}

\subsubsection{Implemented quantitative metric.}

In this study, we aim at quantifying the residual misalignment (noted $T$) between two given images via the Root Mean Square Error criterion (RMSE) computed as follows \cite{TRE}:

\begin{equation}
\mathrm{RMSE}=\frac{1}{\left|\Omega \right|}\sum _{\vec{r} \in \Omega}\sqrt{u(\vec{r})^2 + v(\vec{r})^2 + w(\vec{r})^2}
\end{equation}

\noindent $\vec{r} = (x, y, z)$ being the voxel coordinates, $\Omega$ the image coordinates domain, $\left|\Omega \right|$ the number of voxels in $\Omega$ ($\left|\Omega \right|= 181 \times 217 \times 181$ in the current study) and $T = (u, v, w)$ the voxelwise 3D residual displacement vector field.

The proposed RegQCNET is thus designed to predict registration RMSE using two given images: a reference template and a registered one.

\subsubsection{Implemented deep neural network.} 

\begin{figure}[h!]
\begin{minipage}[b]{\linewidth}
\centering
\centerline{\includegraphics[width=\linewidth]{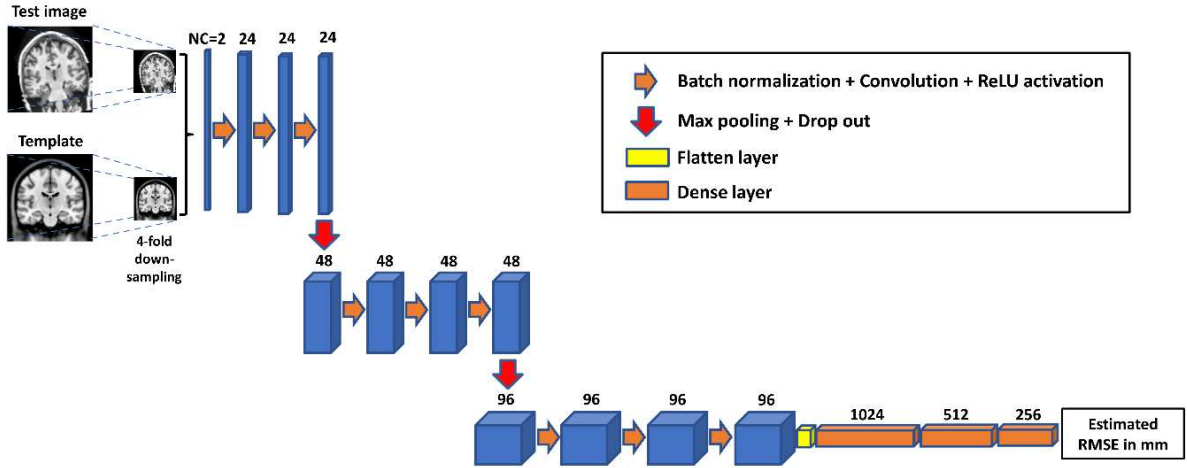}}
\end{minipage}
\caption{Architecture of the proposed RegQCNET for image-to-template registration. Each block is composed of batch normalization, convolution and ReLU activation. The number of input channels (NC) as well as the number of  $3 \times 3 \times 3$ filters are indicated on the top of each block.}
\label{fig:CNN_scheme}
\end{figure}

Figure \ref{fig:CNN_scheme} describes the architecture of the proposed quantitative CNN-based QC for image-to-template registration. Input images were first down-sampled by a factor 4 (note that a down-sampling factor 2 was also tested and discussed). We used a convolutional encoder followed by a 3 regression layers per resolution level using a basis of 24 filters of 3$\times$3$\times$3 (\emph{i.e.}, 24 filters for the first layer, 48 for the second and so on). Each block was composed of batch normalization, convolution and ReLU activation. We employed the following parameters: batch size = 1, optimizer = Adam with default parameters, epoch = 100, loss = Mean Square Error (MSE) and dropout = 0.5 after each block. We used 2 input channels (the down-sampled T1w and the template images).

\subsection{Experimental setup.}

\subsubsection{Assessment of RegQCNET on the ``Simulated testing dataset''.}
\label{sssec:training}

For this dataset, RegQCNET estimations were challenged against real RMSEs by evaluating R$^2$, slope and Y-intercept of a linear regression. While the R$^2$ provides information about the precision of the proposed QC, the slope and Y-intercept quantifies its accuracy. 
We compared results obtained using different sizes for the training dataset (\emph{i.e.}, with $N$=100, 1000 and 10000, respectively). 

To assess the robustness of our method to inaccuracy in image normalization and inhomogeneity correction we performed two experiments. First, we used a given uniform intensity shift to simulate normalization inconsistencies between images. Second, we used a non-uniform intensity bias to simulate inaccurate inhomogeneity correction within images.

\paragraph{Robustness against uniform intensity shift.}

To evaluate the robustness of RegQCNET to incorrect image normalization, RegQCNET was challenged against uniform intensity variation applied on all scans included in the testing dataset. For this purpose, the test scans were identically disturbed as follows: Let $\verb"MAX_INTENSITY"$ be the maximum intensity of an image $I$, and $J$ be the image $I$ after application of the spatially homogeneous intensity bias. For all voxel location $\vec{r}$, the intensity $I(\vec{r})$ was multiplied by a factor 2, while being restricted in the interval original intensity range $\left[0,\right.$\verb"MAX_INTENSITY"$\left.\right]$:

\begin{equation}
J(\vec{r}) = \left\{
    \begin{array}{ll}
         I(\vec{r}) \times 2 & \mbox{if } I(\vec{r}) \times 2 < \verb"MAX_INTENSITY"\\
        \verb"MAX_INTENSITY" & \mbox{otherwise}
    \end{array}
\right.
\end{equation}

\paragraph{Robustness against non-uniform intensity bias.}

To evaluate the robustness of RegQCNET to error during inhomogenety correction, RegQCNET was challenged against a non-uniform intensity bias applied on all scans included in the testing dataset. For this purpose, the test scans were identically disturbed as follows: Let $\vec{r}_0=(x_0,y_0,z_0)$ be the voxel coordinates at the central position of an image $I$, and $K$ be the image $I$ after application of the spatially heterogeneous intensity bias. For all voxel location $\vec{r}$, the intensity $I(\vec{r})$ was weighted by a voxel-wise exponential decay as follow:

\begin{equation}
K(\vec{r}) = I(\vec{r}) \times \exp{\left(-\frac{\| \vec{r} - \vec{r}_0 \|^2 _2 }{2\sigma^2}\right)}
\end{equation}

Practically, intensities in voxels located close to the central position $\vec{r}_0$ were less disturbed than those located further. In the scope of this study, we used $\sigma=60$.

\subsubsection{Assessment of RegQCNET on the ``Real testing dataset''.}
\label{sssec:test2}

First, the 3953 brain images of the ``Real testing dataset'' were visually inspected to build a gold standard. 13 mis-registered brain images were detected and considered as a ``negative case'' for the rest of the manuscript. The ``Simulated training dataset'' with $N=10000$ was here employed for training. The accuracies, area under the ROC curve (AUROC), sensitivity, specificity, positive predictive values (PPVs) and negative predictive values (NPVs) were recorded for the following experiments:

\paragraph{Manually defined threshold.}

RegQCNET was used to differentiate scans with RMSE higher than a user-defined threshold noted $\delta$ ($\delta$ was expressed in millimeters). We tested $\delta$-values in the following set: $\{5, 10, 20, 50\}$ mm. The AUROC was computed using the method detailed in \cite{binary_ROC}.

\paragraph{Automatically defined threshold.}

A 10-fold-stratified cross-validation was used to evaluated the performance of RegQCNET when using an threshold automatically tuned by machine learning algorithm. The dataset of 3953 brains was randomly partitioned into training (90$\%$) and testing (10$\%$) subsets (making sure that at least one positive and one negative case were included in each subset). For this purpose, the following classification algorithms were applied using the commercial software Matlab (\textcopyright1994-2020 The MathWorks, Inc.)/``Statistics and Machine Learning'' toolbox: logistic regression (LR), support vector machine (SVM), na\"ive bayes (NB) and random forest (RF). Default hyper-parameters in Matlab implementations were employed for RF and SVM (RF: Classification method/100 bagged decision trees, SVM: supports sequential minimal optimization/box constraint/linear kernel) \cite{crossvalidation}. The cross-validation steps were repeated 1000 times with shuffling of the folds. Finally, average metrics, standard deviations and confident intervals were calculated.

\paragraph{Comparison with usual image features.}

Our automatically defined threshold experiment was also conducted using correlation coefficient (CC) and mutual information (MI) for comparison.

\subsection{Hardware and implementation details.}

We evaluated the computational burden of our proposed method using an Intel Xeon E5-2683 2.4 GHz (2 Hexadeca-core) with 256 GB of RAM equipped by a GPU Nvidia Tesla V100 with 16 GB of memory. The computation time during the testing session was evaluated without and with the use of the GPU. Our implementation was using Tensorflow 1.4 and Keras 2.2.4.

\section{Results}

\begin{figure}[h!]
\begin{minipage}[b]{0.2\linewidth}
\centering
 \centerline{\small{Template}}\medskip
 \centerline{\small{}}\medskip
 \vspace{1.5cm}
\end{minipage}
\begin{minipage}[b]{0.26\linewidth}
\centering
\centerline{\includegraphics[height=2.4cm]{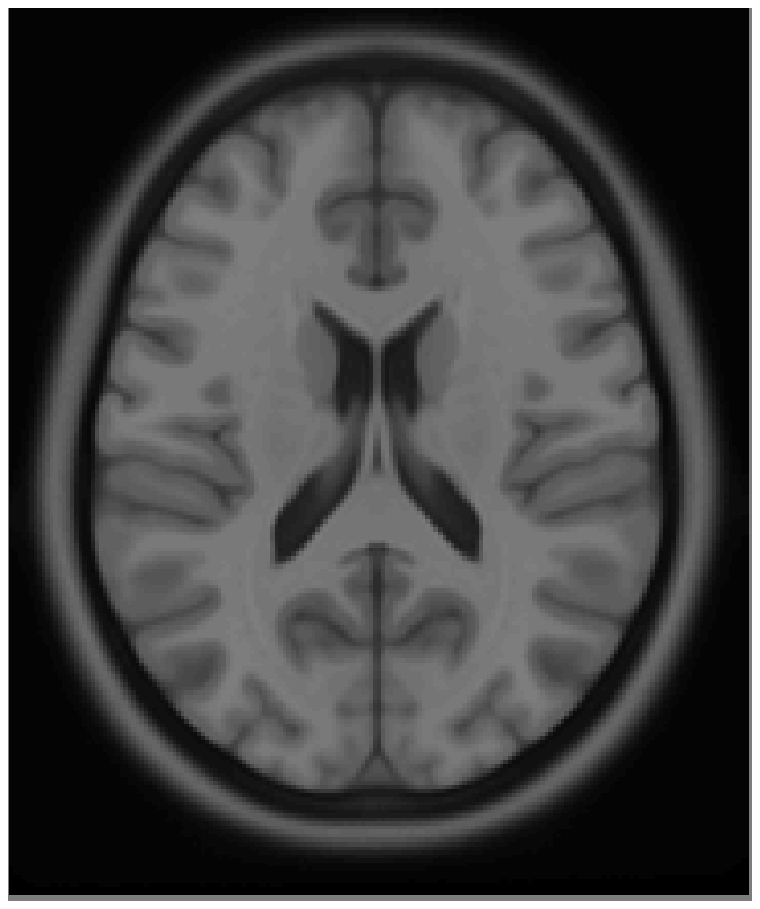}}
\centerline{(a)}\medskip
\end{minipage}
\begin{minipage}[b]{0.26\linewidth}
\centering
\centerline{\includegraphics[height=2.4cm]{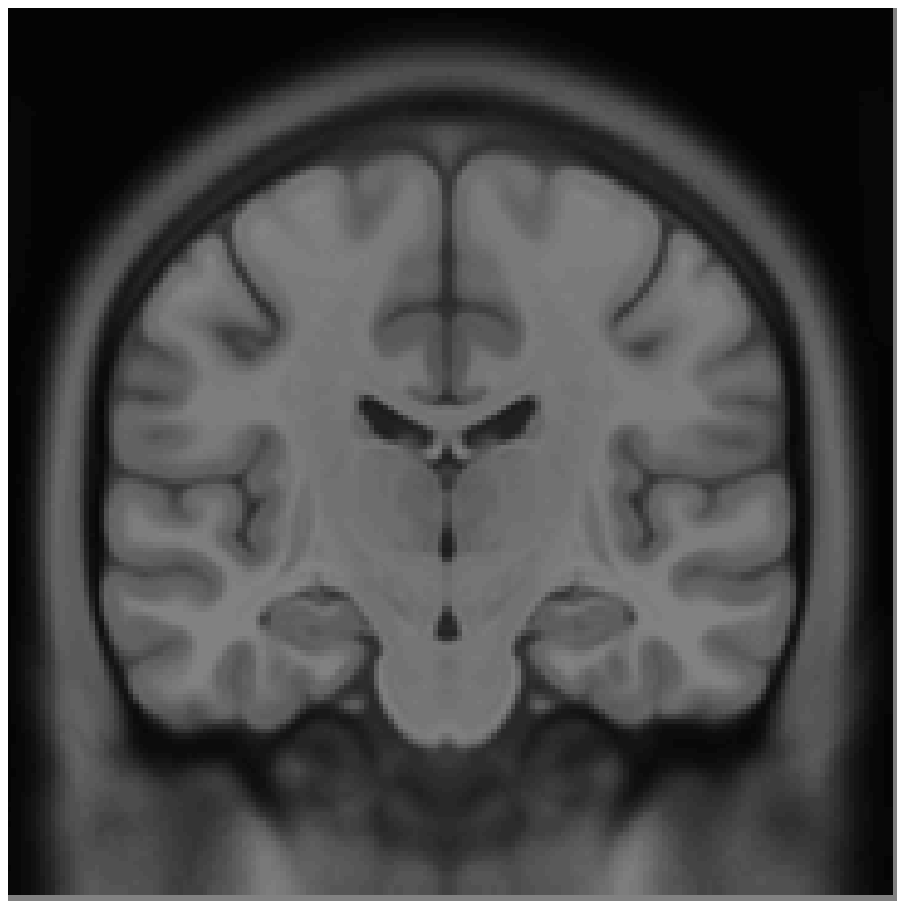}}
\centerline{(b)}\medskip
\end{minipage}
\begin{minipage}[b]{0.26\linewidth}
\centering
\centerline{\includegraphics[height=2.4cm]{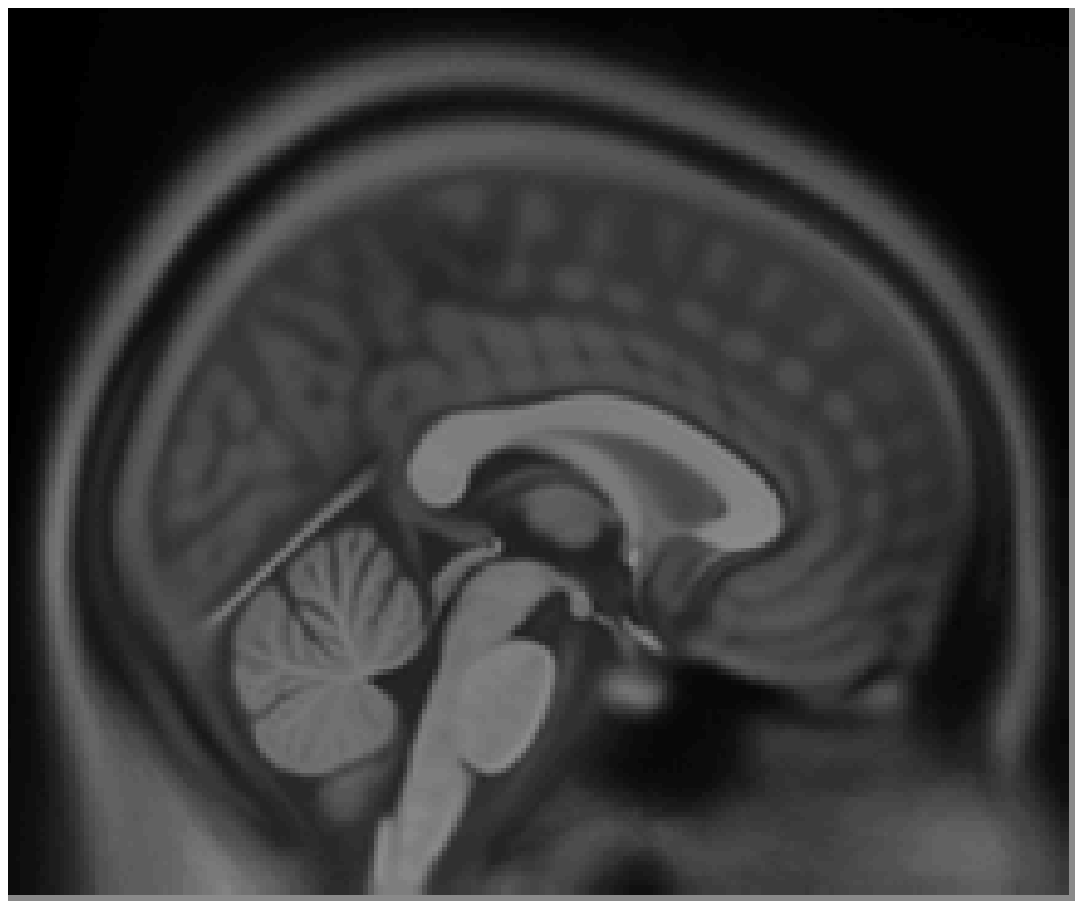}}
\centerline{(c)}\medskip
\end{minipage}
\begin{minipage}[b]{0.2\linewidth}
\centering
 \centerline{\small{Scan $\#1$}}\medskip
 \centerline{\small{RMSE=0}}\medskip
 \vspace{1.5cm}
\end{minipage}
\begin{minipage}[b]{0.26\linewidth}
\centering
\centerline{\includegraphics[height=2.4cm]{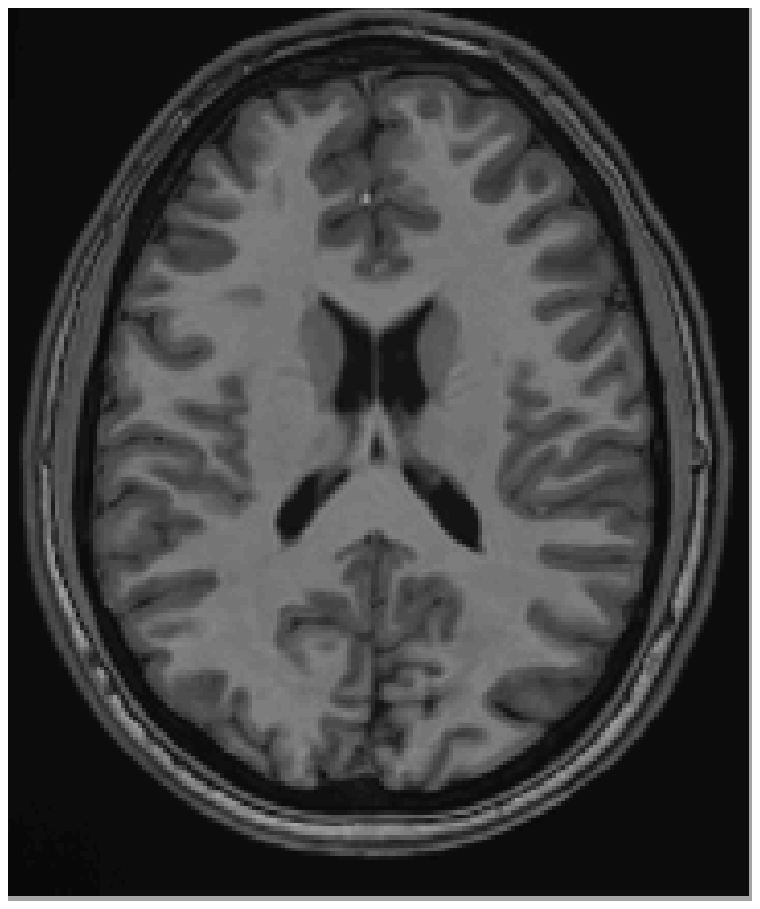}}
\centerline{(d)}\medskip
\end{minipage}
\begin{minipage}[b]{0.26\linewidth}
\centering
\centerline{\includegraphics[height=2.4cm]{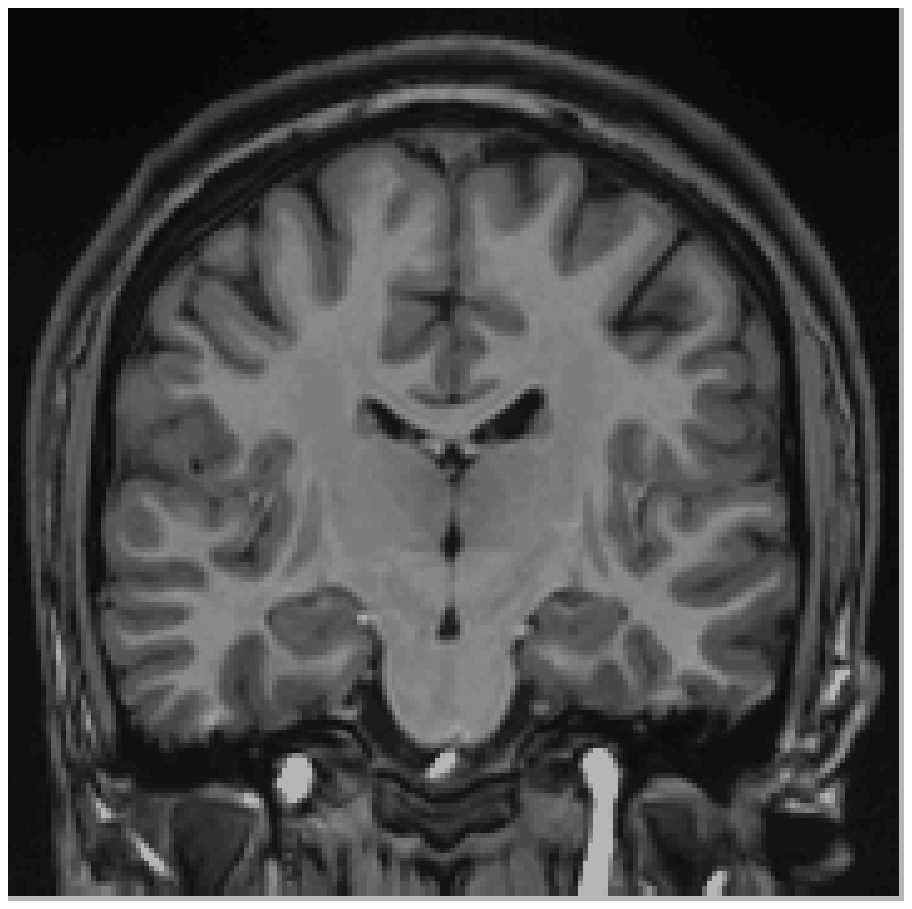}}
\centerline{(e)}\medskip
\end{minipage}
\begin{minipage}[b]{0.26\linewidth}
\centering
\centerline{\includegraphics[height=2.4cm]{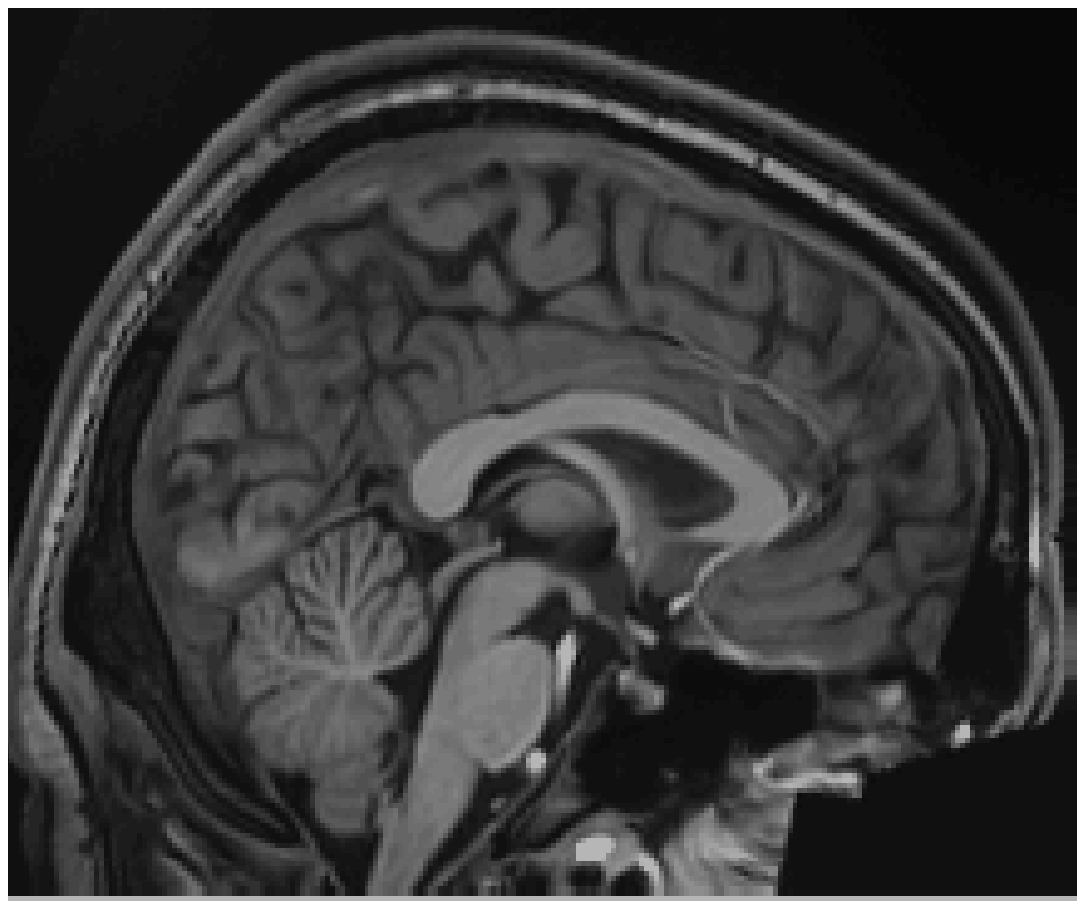}}
\centerline{(f)}\medskip
\end{minipage}
\begin{minipage}[b]{0.2\linewidth}
\centering
 \centerline{\small{Scan $\#2$}}\medskip
 \centerline{\small{RMSE=10}}\medskip
 \vspace{1.5cm}
\end{minipage}
\begin{minipage}[b]{0.26\linewidth}
\centering
\centerline{\includegraphics[height=2.4cm]{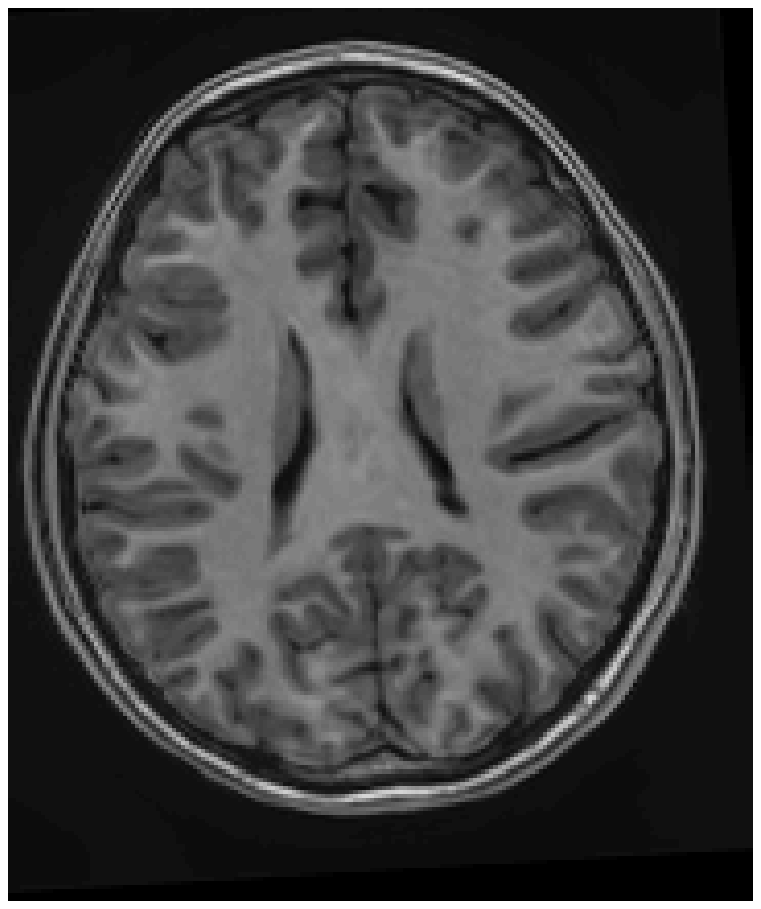}}
\centerline{(g)}\medskip
\end{minipage}
\begin{minipage}[b]{0.26\linewidth}
\centering
\centerline{\includegraphics[height=2.4cm]{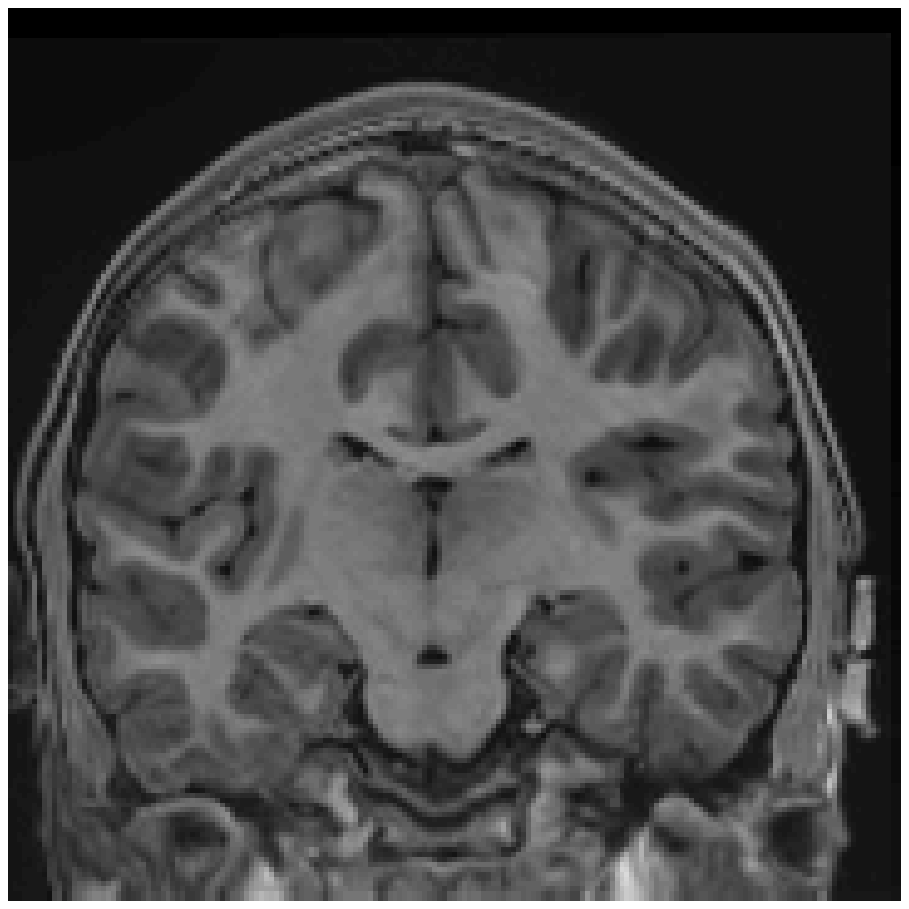}}
\centerline{(h)}\medskip
\end{minipage}
\begin{minipage}[b]{0.26\linewidth}
\centering
\centerline{\includegraphics[height=2.4cm]{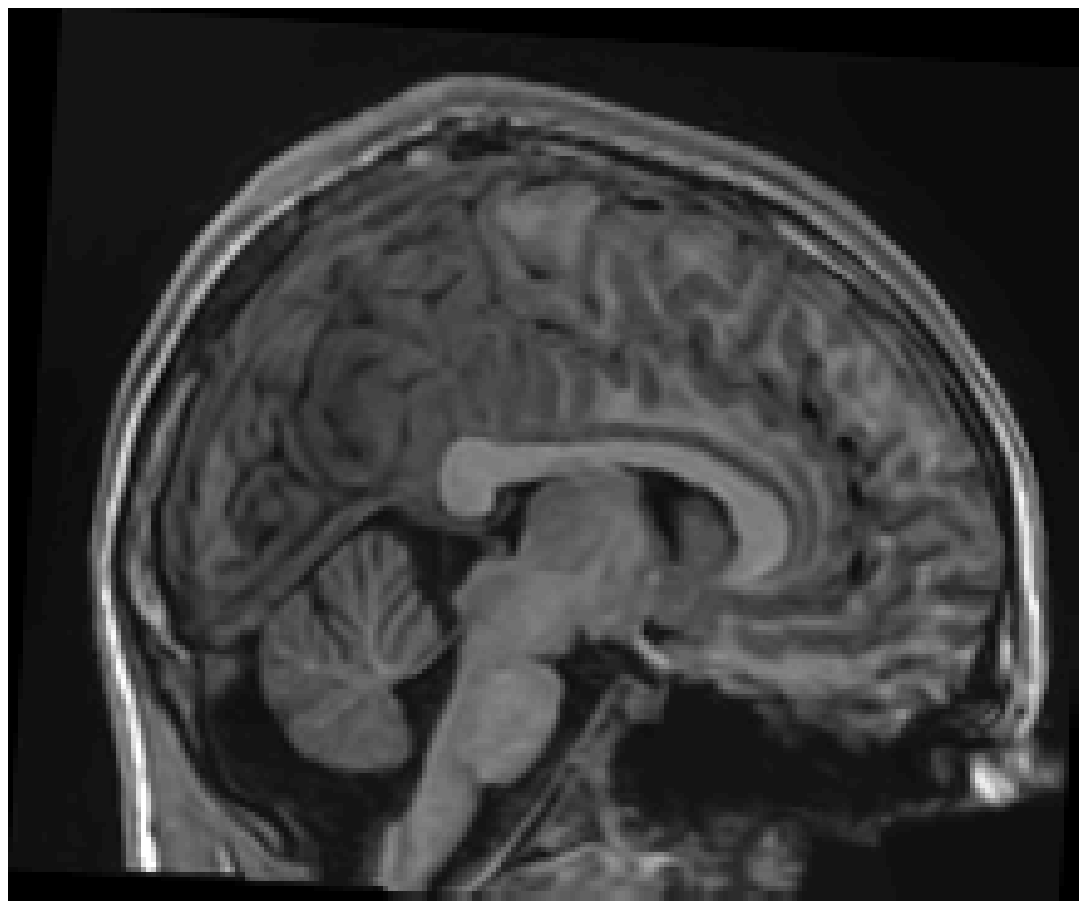}}
\centerline{(i)}\medskip
\end{minipage}
\begin{minipage}[b]{0.2\linewidth}
\centering
 \centerline{\small{Scan $\#3$}}\medskip
 \centerline{\small{RMSE=21}}\medskip
 \vspace{1.5cm}
\end{minipage}
\begin{minipage}[b]{0.26\linewidth}
\centering
\centerline{\includegraphics[height=2.4cm]{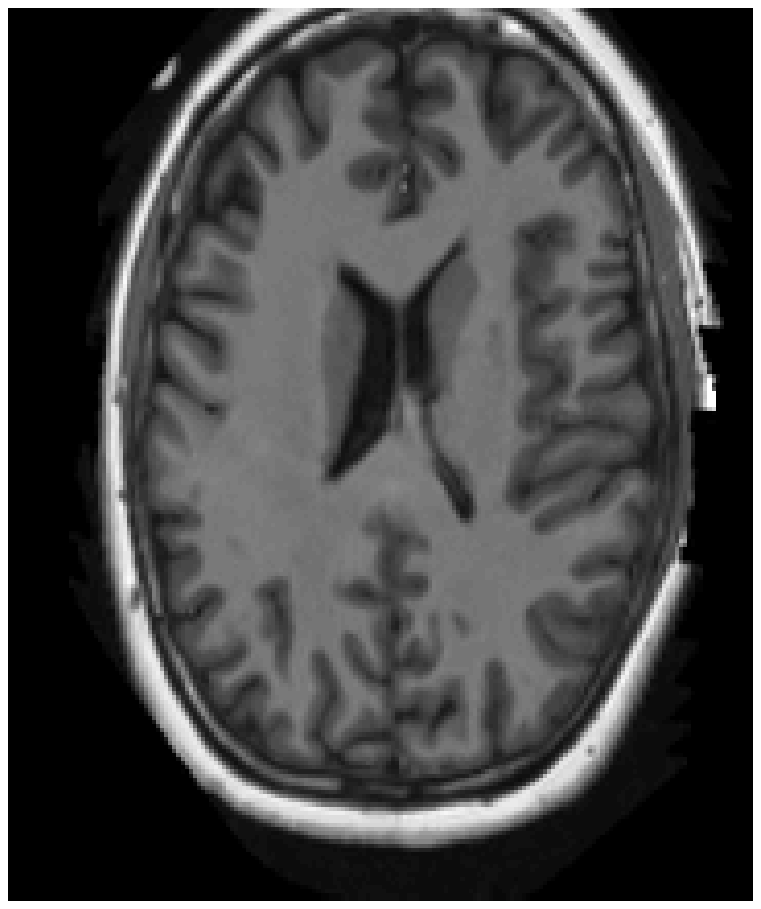}}
\centerline{(j)}\medskip
\end{minipage}
\begin{minipage}[b]{0.26\linewidth}
\centering
\centerline{\includegraphics[height=2.4cm]{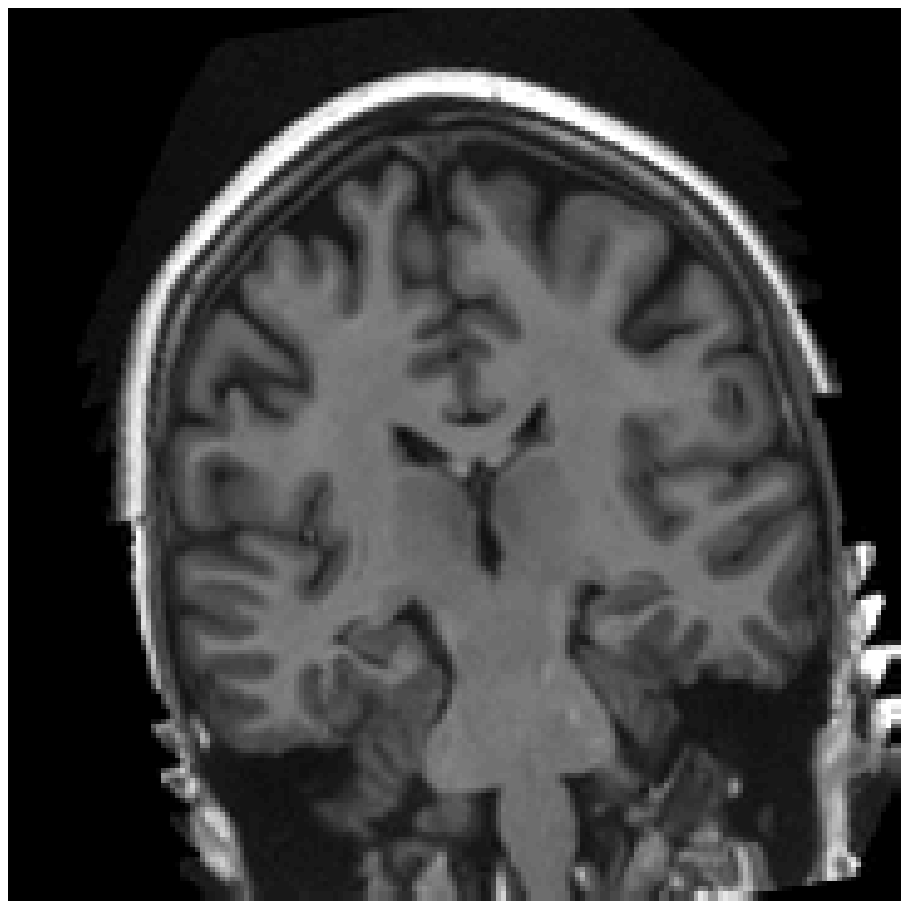}}
\centerline{(k)}\medskip
\end{minipage}
\begin{minipage}[b]{0.26\linewidth}
\centering
\centerline{\includegraphics[height=2.4cm]{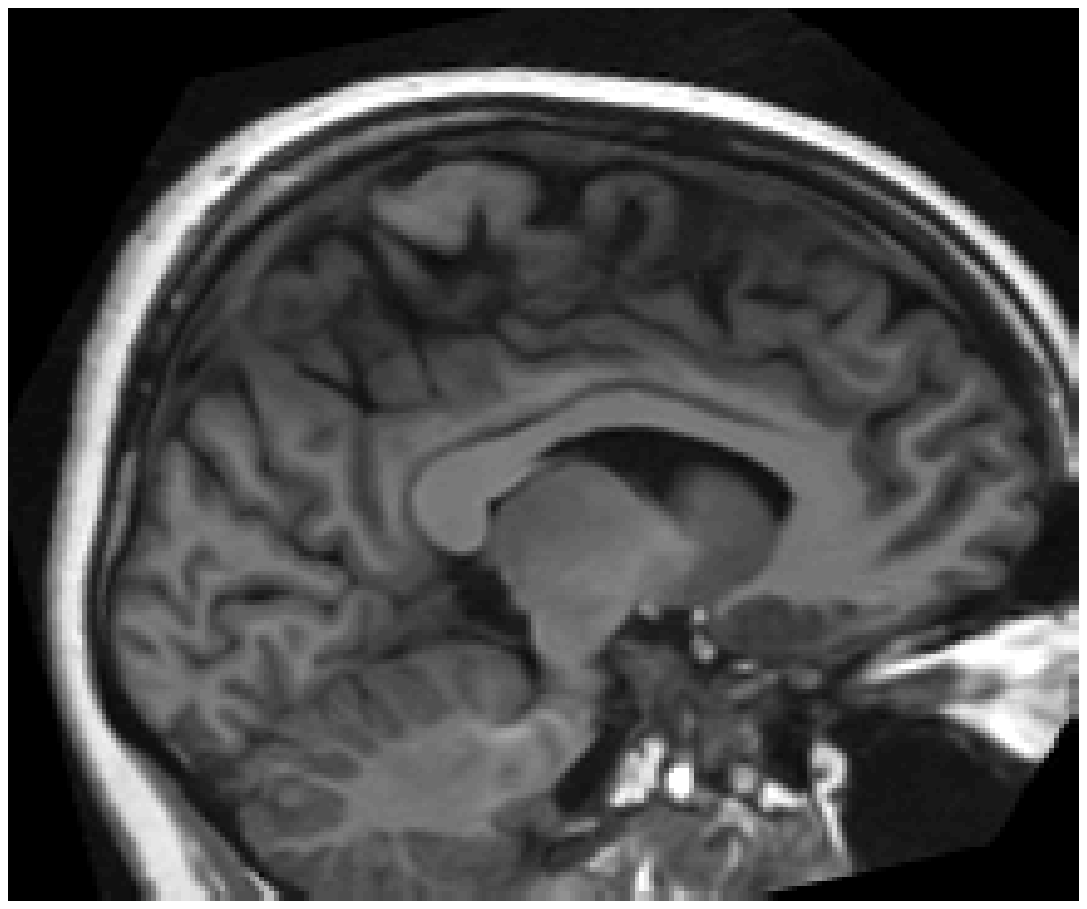}}
\centerline{(l)}\medskip
\end{minipage}
\begin{minipage}[b]{0.2\linewidth}
\centering
 \centerline{\small{Scan $\#4$}}\medskip
 \centerline{\small{RMSE=41}}\medskip
 \vspace{1.5cm}
\end{minipage}
\begin{minipage}[b]{0.26\linewidth}
\centering
\centerline{\includegraphics[height=2.4cm]{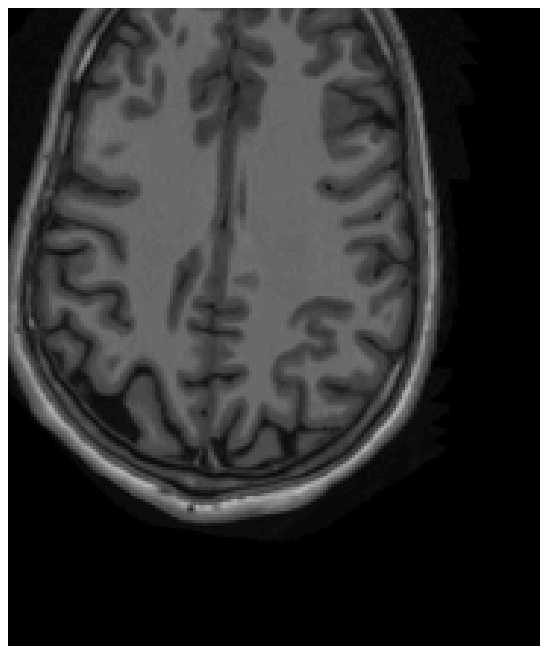}}
\centerline{(m)}\medskip
\end{minipage}
\begin{minipage}[b]{0.26\linewidth}
\centering
\centerline{\includegraphics[height=2.4cm]{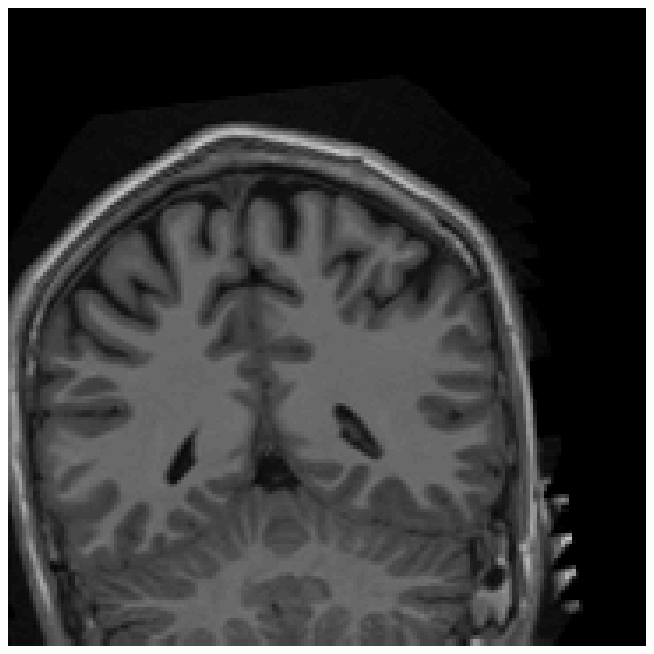}}
\centerline{(n)}\medskip
\end{minipage}
\begin{minipage}[b]{0.26\linewidth}
\centering
\centerline{\includegraphics[height=2.4cm]{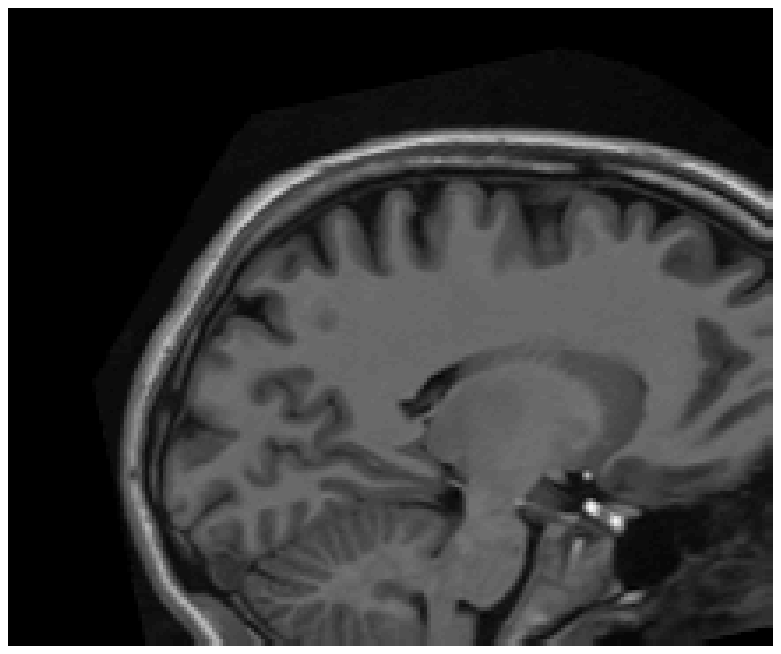}}
\centerline{(o)}\medskip
\end{minipage}
\begin{minipage}[b]{0.2\linewidth}
\centering
 \centerline{\small{Scan $\#5$}}\medskip
 \centerline{\small{RMSE=99}}\medskip
 \vspace{1.5cm}
\end{minipage}
\begin{minipage}[b]{0.26\linewidth}
\centering
\centerline{\includegraphics[height=2.4cm]{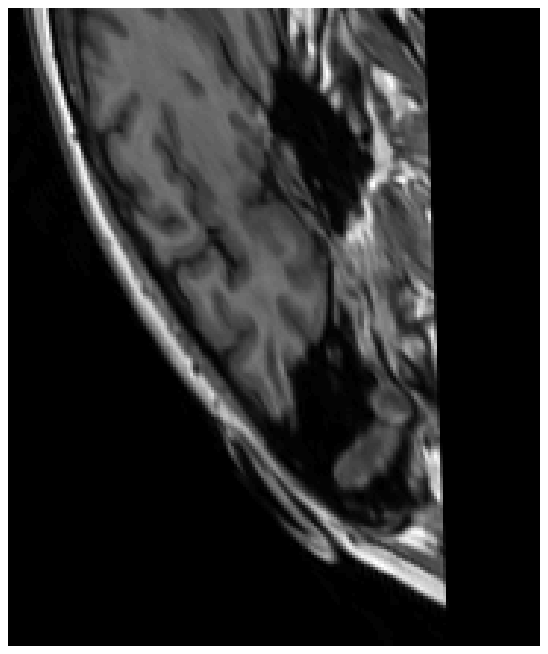}}
\centerline{(p)}\medskip
\end{minipage}
\begin{minipage}[b]{0.26\linewidth}
\centering
\centerline{\includegraphics[height=2.4cm]{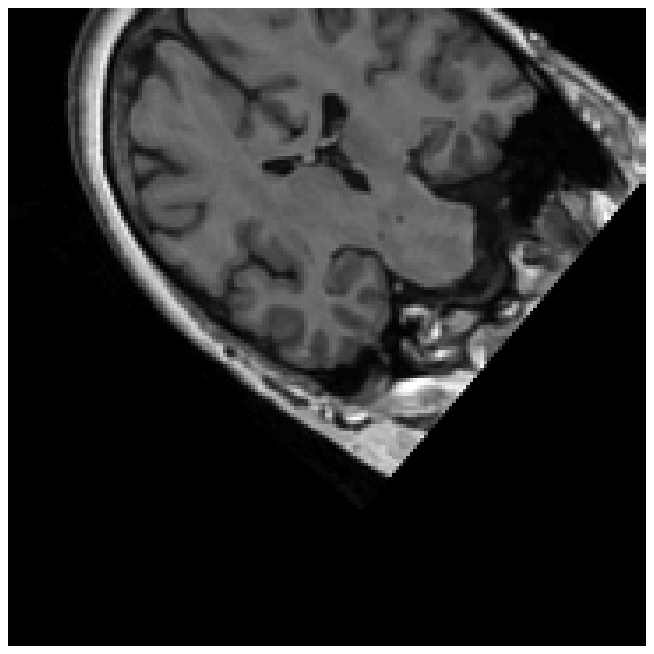}}
\centerline{(q)}\medskip
\end{minipage}
\begin{minipage}[b]{0.26\linewidth}
\centering
\centerline{\includegraphics[height=2.4cm]{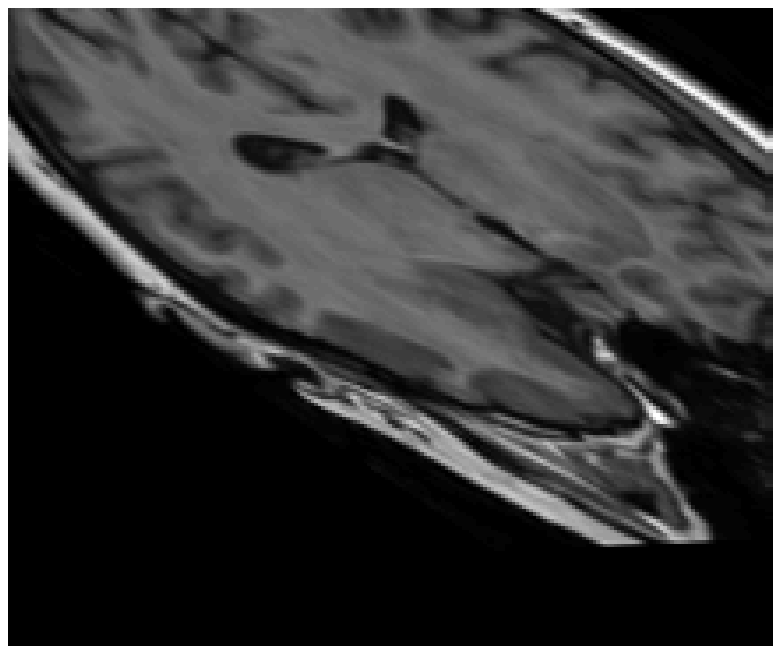}}
\centerline{(r)}\medskip
\end{minipage}
\caption{Typical images generated using synthetic affine transformations. Transversal (left column), coronal (middle column) and sagittal (right column) central slice are reported for the template scan (a-c) and for different subjects/various RMSE values. Each generated image is referred to as ``Scan $\#[1-5]$''.}
\label{fig:scans}
\end{figure}

Fig. \ref{fig:scans} shows typical images generated using synthetic affine transformations. Middle transversal, coronal and sagittal slices are reported for several 3D volumes. The template used as reference for affine image registration (see section \ref{sssec:preprocessing}) is displayed in the first row. The second (scan $\#1$) row shows 3D brain images from the original lifespan dataset (RMSE considered equal to 0 mm). Lower rows (scan $\#[3-5]$) show examples of 3D scans obtained after the application of simulated spatial transformations of various amplitudes, as described in section \ref{sssec:trans}. Note that several images underwent different masking (see (f) on the second row) due to defacing (e.g., NDAR dataset).

\subsection{Assessment of RegQCNET on the ``Simulated testing dataset''.}

Fig. \ref{fig:RMSE} reports the precision and the accuracy of RegQCNET obtained on the ``simulated testing dataset'' (as described in section \ref{sssec:test1}) using a training dataset of 100 (\ref{fig:RMSE}a), 1000 (\ref{fig:RMSE}b) and 10000 (\ref{fig:RMSE}c) scans. As one can expect, the precision (rated by the R$^2$ of the linear fit) improves when the size of the training dataset increased. The R$^2$ converged slowly toward 1 along with the size of the training dataset increased (R$^2$ equal to 0.84, 0.95 and 0.99 were obtained using 100, 1000 and 10000 images, respectively). The accuracy (rated by the slope and the Y-intercept of the linear regression) followed the same trend. As long as $N$ increased, the slope and the Y-intercept converged toward optimal values (\emph{i.e.}, 1 and 0, respectively). 

Fig. \ref{fig:intensity_bias} shows the impact uniform intensity shift and spatially varying intensity bias (as described in section \ref{sssec:test1}) on the performance of the proposed RegQCNET. While a uniform intensity shift did not deteriorate the precision (R$^2$=0.99) and the accuracy (slope=0.98/Y-intercept=-1.08) (\ref{fig:intensity_bias}c) this observation did not hold when a non-uniform bias (\ref{fig:intensity_bias}e): in this last case, both precision (R$^2$=0.98) and accuracy (slope=0.89/Y-intercept=7.54) were slightly worse.

\subsection{Assessment of RegQCNET on the ``Real testing dataset''.}

\begin{figure}[h!]
\begin{minipage}[b]{\linewidth}
\centering
\centerline{\includegraphics[width=\linewidth]{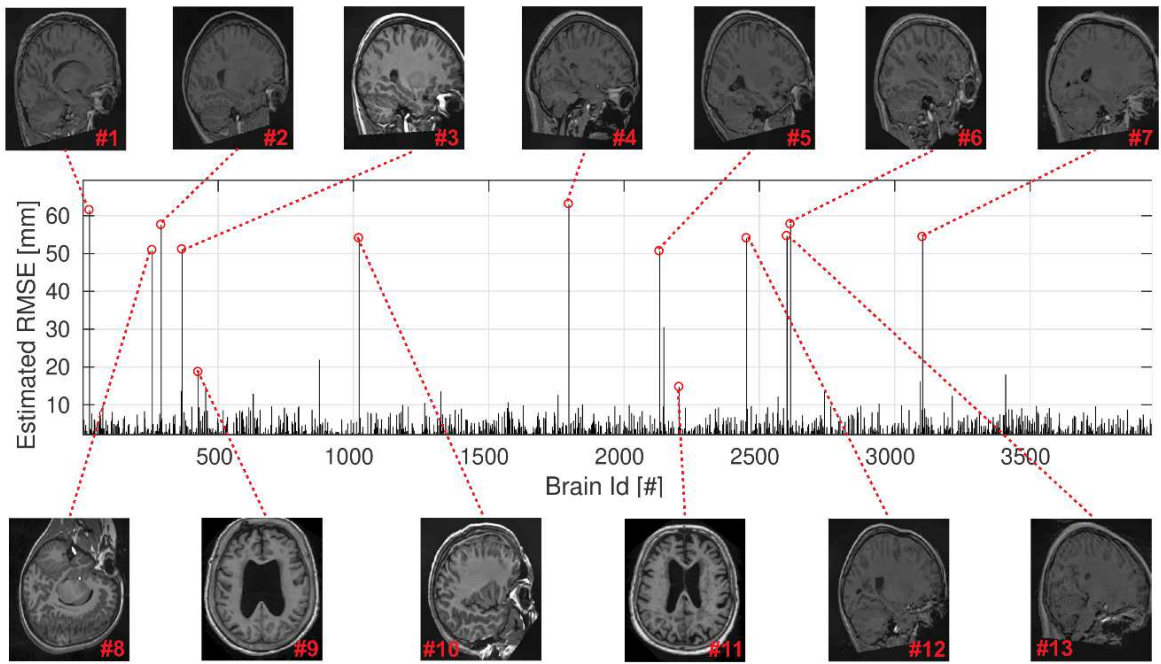}}
\end{minipage}
\caption{RMSEs estimated by RegQCNET on the 3953 brains of the data base. Mis-registered images (transversal), as detected by visual inspection, are reported above and below the graph. The corresponding template cross-section is shown in Fig. \ref{fig:scans}a.}
\label{fig:realwordscenario}
\end{figure}

Fig. \ref{fig:realwordscenario} shows RMSE estimated by RegQCNET on the 3953 tested brain MRIs. Transversal cross-section of mis-registered images, as detected by visual inspection, are reported above and below the graph. These images can be visually compared to the corresponding template cross-section reported in Fig. \ref{fig:scans}a. In two images (referred to as case $\#9$ and case $\#11$ in Fig. \ref{fig:realwordscenario}) of patients with Alzeihmer's Disease, very large lateral ventricle slightly disturbed RegQCNET (estimated RMSE $<20$ mm). In the 9 other images, huge mis-registrations are observable, which have been detected by RegQCNET (estimated RMSE $>50$ mm). 

\subsubsection{Manually defined threshold.}

Table \ref{table:manual_classification} reports classification scores of RegQCNET using different manually defined thresholds. A classification threshold $\delta=10$ mm provided best scores: accuracy=99.6$\%$, AUROC=1.0, sensitivity=99.6$\%$, specificity=100.0$\%$, PPV=100.0$\%$, and NPV=44.8$\%$ (note that NPV=44.8$\%$ means here that 16 good-registered images were considered as mis-registered). Good- and mis-registered images were thus correctly identified in 3924/3940 and 13/13 brain images, respectively.

\subsubsection{Automatically defined threshold.}

The RegQCNET output served as a metric in all tested machine learning classifiers (see Table \ref{table:classification}). In particular, using a logistic regression classifier, QC scores were: accuracy=96.0$\%$, AUROC=1.0, sensitivity=95.9$\%$, specificity=100.0$\%$, PPV=97.5$\%$, and NPV=93.7$\%$. %In average over the 1000 cross-validation repetitions, good- and mis-registered images were thus correctly identified in 3939/3940 and 13/13 brain images, respectively. 

\subsubsection{Comparison with usual image metrics.}

\begin{figure}[h!]
\begin{minipage}[b]{0.32\linewidth}
\centerline{\hspace{0.3cm}RegQCNET / LR}\medskip
\centerline{\includegraphics[trim={0cm 0cm 0cm 0cm},clip,height=5cm]
{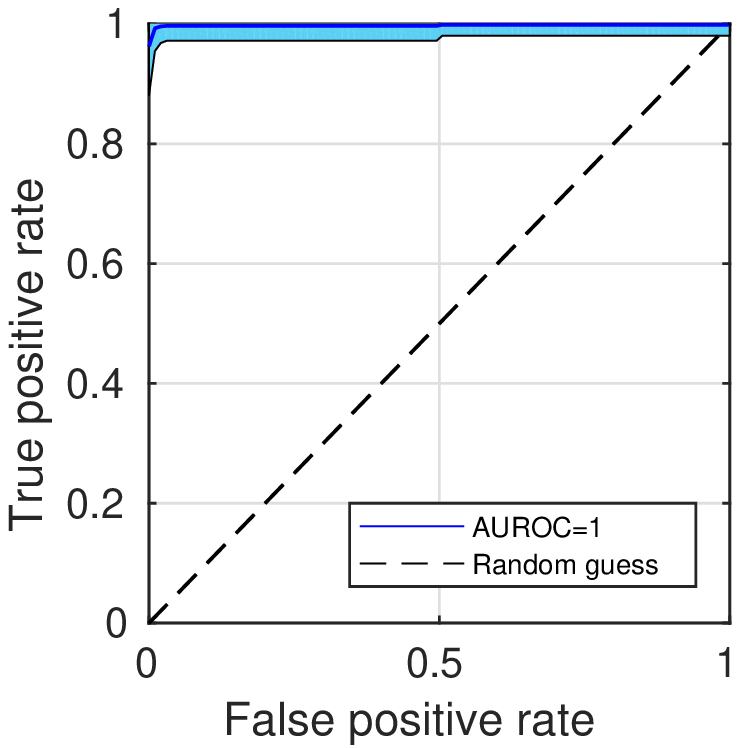}}
\centerline{(a)}\medskip
\end{minipage}
\begin{minipage}[b]{0.32\linewidth}
\centerline{\hspace{0.3cm}Cross-correlation/ LR}\medskip
\centerline{\includegraphics[trim={0cm 0cm 0cm 0cm},clip,height=5cm]
{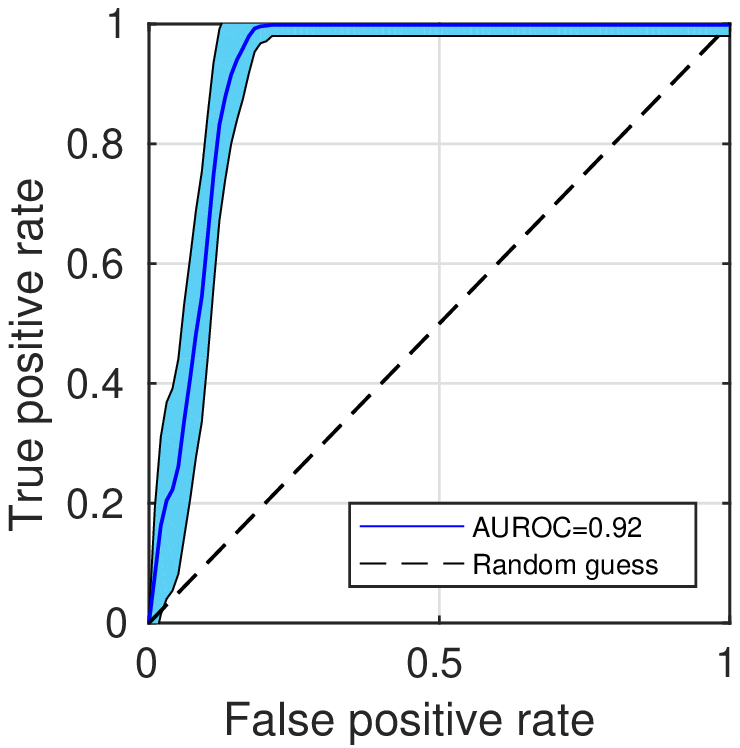}}
\centerline{(b)}\medskip
\end{minipage}
\begin{minipage}[b]{0.32\linewidth}
\centerline{\hspace{0.3cm}Mutual information / NB}\medskip
\centerline{\includegraphics[trim={0cm 0cm 0cm 0cm},clip,height=5cm]
{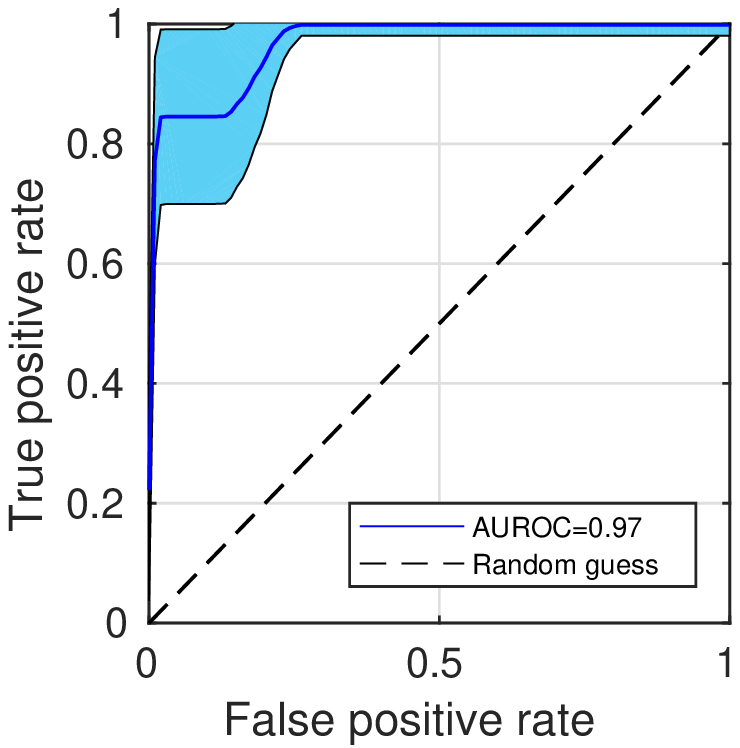}}
\centerline{(c)}\medskip
\end{minipage}
\caption{ROC curves obtained using the three tested indicators (RegQCNET (a), CC (b) and MI (c)) as binary classifiers (LR (a), LR (b) and NB (c)) for the two image populations (\emph{i.e.}, correctly registered vs. mis-registered) after 10-fold cross-validation.}
\label{fig:ROCcurves}
\end{figure}

Very poor scores were obtained using CC (best classifier=na\"ive bayes) and MI (best classifier=logistic regression): a good detection of correctly registered images was achievable by accepting a large amount of false-negative cases, as shown in Fig. \ref{fig:ROCcurves}. Conversely, a perfect detection of mis-registered images was only achievable by accepting a dramatic impact on the sensitivity (\emph{i.e.}, 1.7$\%$ and 30.9$\%$ for CC and MI, respectively, as shown in Table \ref{table:classification}).

\begin{figure}[h!]
\begin{minipage}[b]{0.49\linewidth}
\centerline{\hspace{0.7cm}100 training scans}\medskip
\centerline{\includegraphics[trim={0cm 0cm 0cm 0cm},clip,height=6cm]
{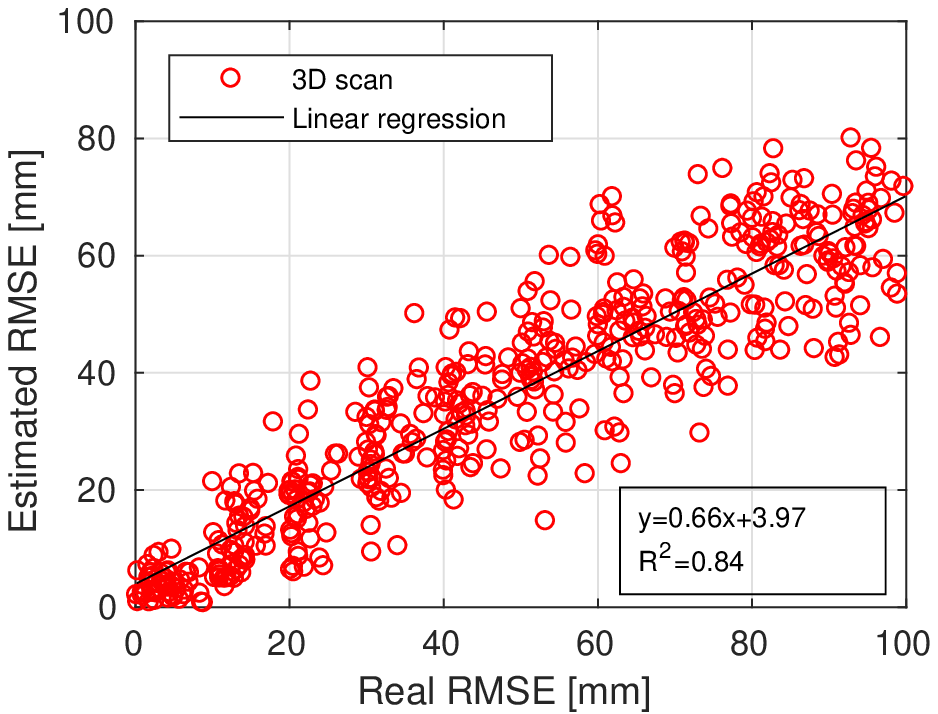}}
\centerline{(a)}\medskip
\end{minipage}
\begin{minipage}[b]{0.49\linewidth}
\centerline{\hspace{0.7cm}1000 training scans}\medskip
\centerline{\includegraphics[trim={0cm 0cm 0cm 0cm},clip,height=6cm]
{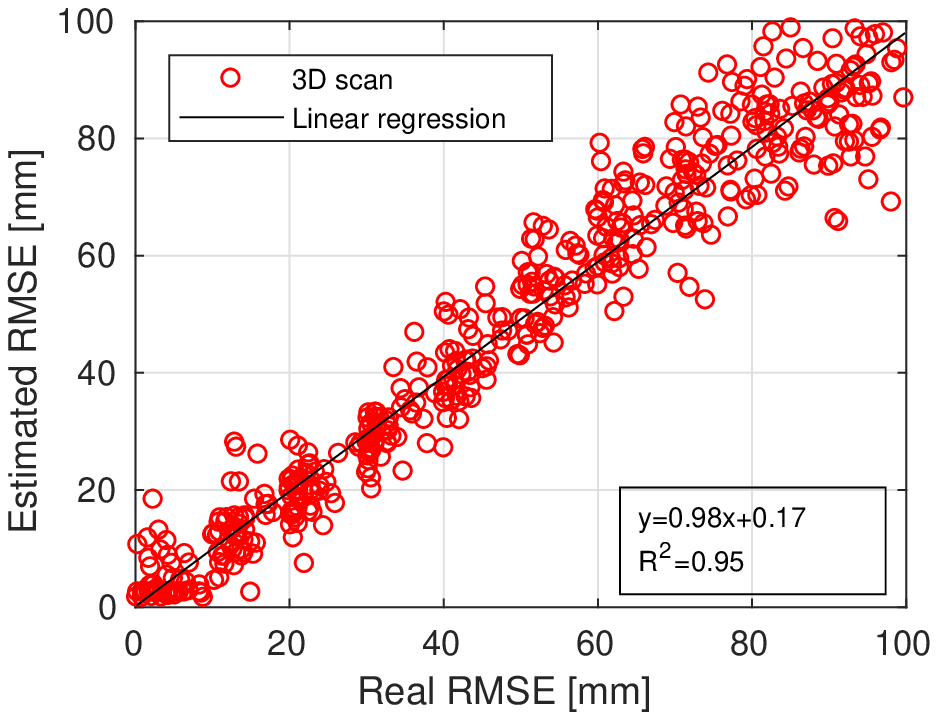}}
\centerline{(b)}\medskip
\end{minipage}
\begin{minipage}[b]{\linewidth}
\centerline{\hspace{0.7cm}10000 training scans}\medskip
\centerline{\includegraphics[trim={0cm 0cm 0cm 0cm},clip,height=6cm]
{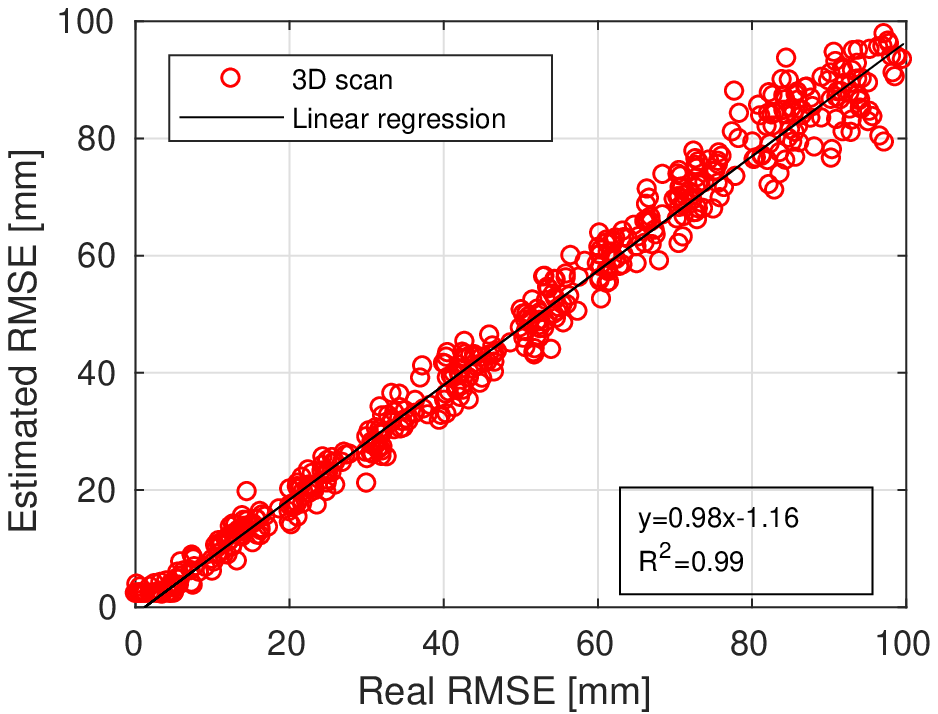}}
\centerline{(c)}\medskip
\end{minipage}
\caption{RMSE estimated by RegQCNET on the simulated dataset using training set composed of 100 (a),1000 (b) and 10000 (c) scans. Estimated RMSEs are plotted against real RMSEs and the R$^2$, slope and Y-intercept of a linear regression are reported in the insert of each graph.}
\label{fig:RMSE}
\end{figure}

\begin{figure}[h!]
\begin{minipage}[b]{0.25\linewidth}
\centering
\vspace{2.5cm}
\centerline{\small{Original}}\medskip
\centerline{\includegraphics[trim={0.1cm 0cm 0cm 0.0cm},clip,height=4cm]
{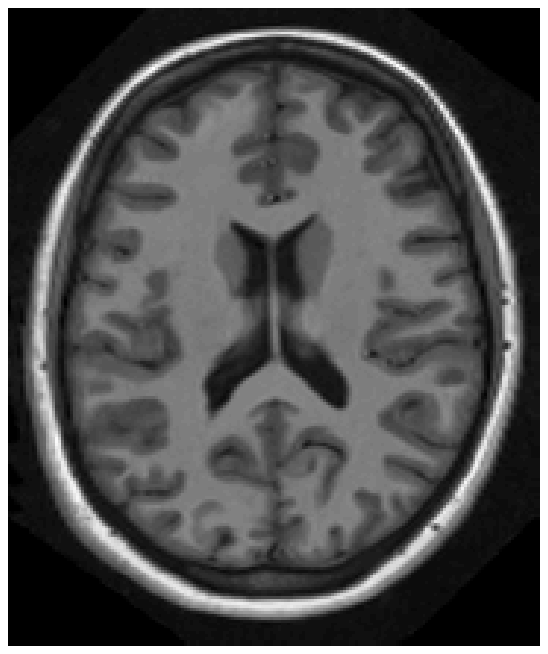}}
\centerline{(a)}\medskip
\vspace{-2.5cm}
\end{minipage}
\begin{minipage}[b]{0.25\linewidth}
\centering
%\vspace{-1.5cm}
\centerline{\small{Uniform}}
\centerline{\small{intensity shift}}\medskip
\centerline{\includegraphics[trim={0.1cm 0cm 0cm 0.0cm},clip,height=4cm]
{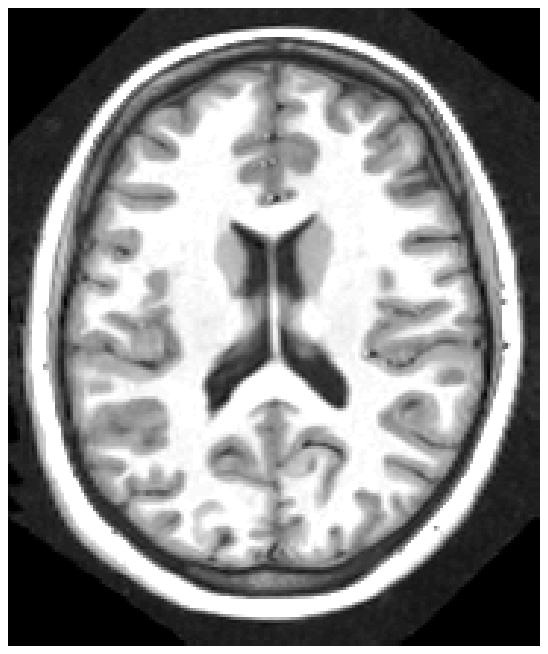}}
\vspace{0.5cm}
\centerline{(b)}\medskip

\end{minipage}
\begin{minipage}[b]{0.49\linewidth}
\centering
\centerline{\includegraphics[trim={0cm 0cm 0cm 0cm},clip,height=5.5cm]
{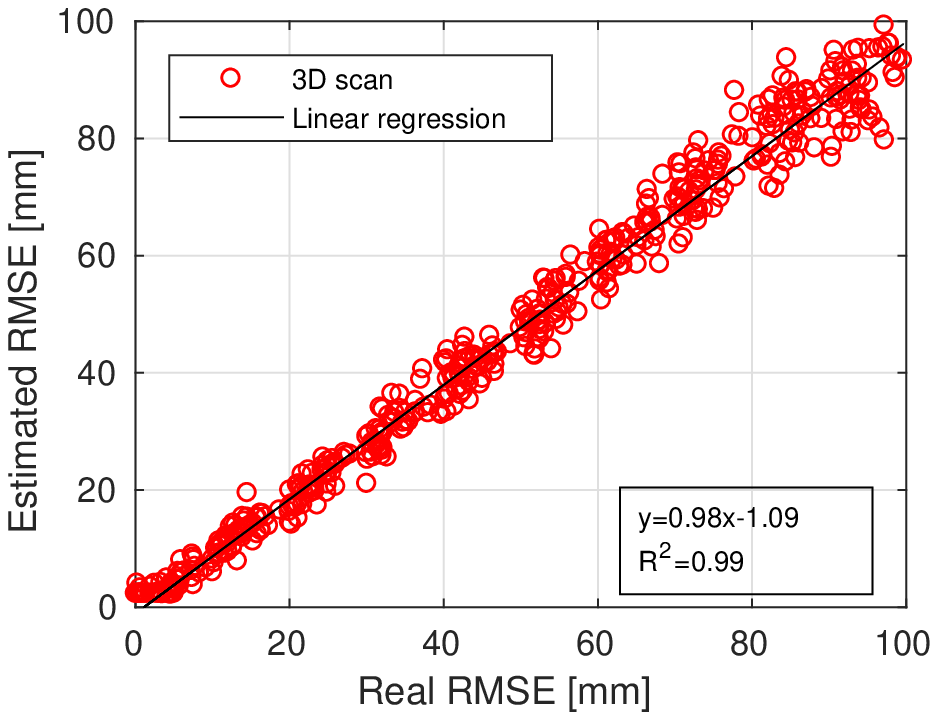}}
\centerline{(c)}\medskip
\end{minipage}

\begin{minipage}[b]{0.25\linewidth}
\centering
\centerline{ }\medskip
\end{minipage}
\begin{minipage}[b]{0.25\linewidth}
\centering
\centerline{\small{Non-uniform}}
\centerline{\small{intensity bias}}\medskip
\centerline{\includegraphics[trim={0.1cm 0cm 0cm 0.0cm},clip,height=4cm]
{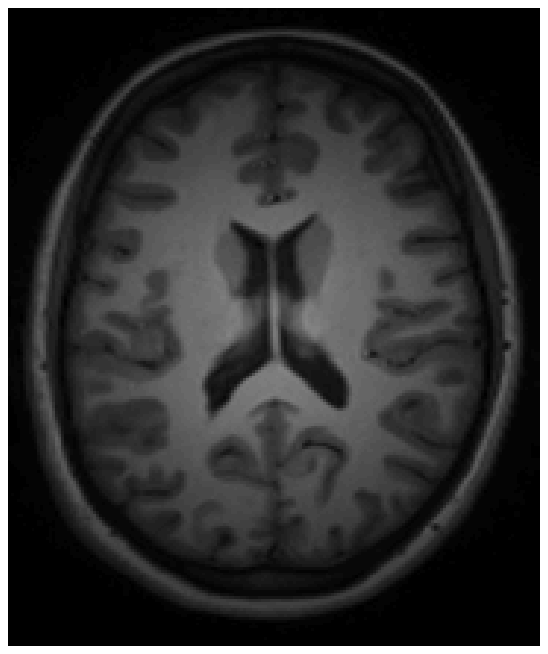}}
\vspace{0.5cm}
\centerline{(d)}\medskip
\end{minipage}
\begin{minipage}[b]{0.49\linewidth}
\centering
\centerline{\includegraphics[trim={0cm 0cm 0cm 0cm},clip,height=5.5cm]
{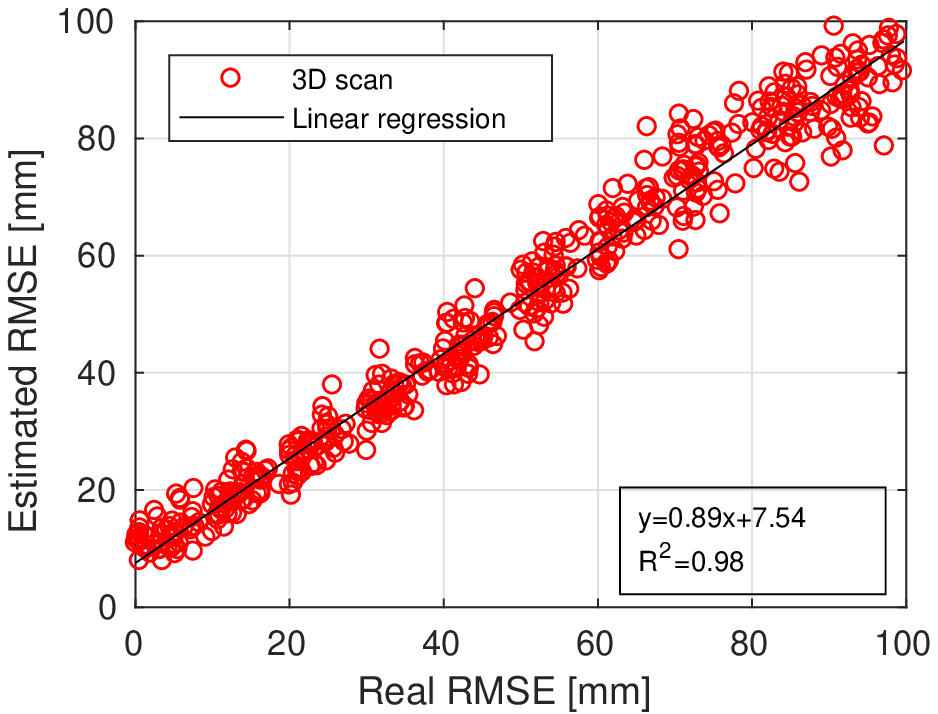}}
\centerline{(e)}\medskip
\end{minipage}
\caption{Results obtained on the simulation experiment when an intensity perturbation is applied on the testing dataset. A training dataset composed by 10000 scans (without intensity bias) was used. The axial slice of a brain scan is reported before (a) and after application of an uniform (b) and a non-uniform (d) intensity bias. Estimated RMSEs are plotted against real RMSEs for the uniform (c) and the non-uniform (e) bias, and the R$^2$, slope and Y-intercept of a linear regression are reported.}
\label{fig:intensity_bias}
\end{figure}

\begin{table}
\begin{tabular*}{\textwidth}{@{}c*{6}{cccccc}}
 \hline
% \vspace{-0.2cm}
\multicolumn{7}{c}{Classification scores: Manually defined threshold}\\
%\vspace{-0.3cm}
\cline{1-7}\\
Classification&Accuracy&AUROC&Sensitivity&Specificity&PPV&NPV\\
threshold ($\delta$) [mm]&&&&&&\\
\hline
5 & 91.6 & 0.96 & 91.6 & 100.0 & 100.0 & 3.8\\
\bf{10} & \bf{99.6} & \bf{1.00} & \bf{99.6} & \bf{100.0} & \bf{100.0} & \bf{44.8}\\
20 & 99.9 & 0.92 & 99.9 & 84.6 & 99.9 & 84.6\\
50 & 99.9 & 0.92 & 100.0 & 84.6 & 99.9 & 100.0\\
\hline
\end{tabular*}
\caption{Classification scores of the proposed RegQCNET on the ``Real testing dataset''. Quantitative scores (\emph{i.e.}, computer-assisted) are given, assuming the qualitative inspection (\emph{i.e.}, visual) as a gold standard. AUROC: area under the ROC curve; PPV: positive predictive value; NPV: negative predictive value. Accuracies, sensitivities, specificities, PPVs, and NPVs are shown in percentages. Best performance are reported in bold font.}
\label{table:manual_classification}
\end{table}

\begin{table}
\begin{tabular*}{\textwidth}{@{}l*{6}{ccccc}}
 \hline
% \vspace{-0.2cm}
& \multicolumn{5}{c}{Classification scores: Automatically defined threshold}\\
%\vspace{-0.3cm}
&\cline{1-5}\\
Classifier&Accuracy&AUROC&Sensitivity&Specificity&PPV&NPV\\
\hline
\multicolumn{2}{l}{\hspace{-0.2cm}\small{RegQCNET}}& & & & & & & \\
\hspace{0.5cm}\bf{\footnotesize{LR}}&\bf{\footnotesize{96.0$\pm$17.3}}&\bf{\footnotesize{1.00$\pm$0.06}}&\bf{\footnotesize{95.9$\pm$17.4}}&\bf{\footnotesize{100.0$\pm$0.0}}&\bf{\footnotesize{97.5$\pm$15.7}}&\bf{\footnotesize{93.7$\pm$24.2}}\\
&\bf{\footnotesize{(94.7-97.2)}}&\bf{\footnotesize{(0.99-1.00)}}&\bf{\footnotesize{(94.7-97.2)}}&\bf{\footnotesize{(100.0-100.0)}}&\bf{\footnotesize{(96.4-98.6)}}&\bf{\footnotesize{(92.0-95.4)}}\\
\hspace{0.5cm}\footnotesize{SVM}&\footnotesize{94.8$\pm$18.8}&\footnotesize{1.00$\pm$0.04}&\footnotesize{94.8$\pm$18.9}&\footnotesize{100.0$\pm$0.0}&\footnotesize{97.3$\pm$16.2}&\footnotesize{91.3$\pm$28.1}\\
&\footnotesize{(93.4-96.1)}&\footnotesize{(1.00-1.00)}&\footnotesize{(93.4-96.1)}&\footnotesize{(100.0-100.0)}&\footnotesize{(96.1-98.5)}&\footnotesize{(89.3-93.3)}\\
\hspace{0.5cm}\footnotesize{NB}&\footnotesize{85.7$\pm$34.1}&\footnotesize{1.00$\pm$0.04}&\footnotesize{85.6$\pm$34.2}&\footnotesize{100.0$\pm$0.0}&\footnotesize{86.8$\pm$33.9}&\footnotesize{84.0$\pm$36.6}\\
&\footnotesize{(83.3-88.1)}&\footnotesize{(1.00-1.00)}&\footnotesize{(83.2-88.1)}&\footnotesize{(100.0-100.0)}&\footnotesize{(84.4-89.2)}&\footnotesize{(81.4-86.6)}\\
\hspace{0.5cm}\footnotesize{RF}&\footnotesize{84.8$\pm$30.7}&\footnotesize{0.94$\pm$0.18}&\footnotesize{84.8$\pm$30.8}&\footnotesize{100.0$\pm$0.0}&\footnotesize{91.3$\pm$28.5}&\footnotesize{76.2$\pm$42.9}\\
&\footnotesize{(75.7-93.9)}&\footnotesize{(0.89-1.00)}&\footnotesize{(75.6-93.9)}&\footnotesize{(100.0-100.0)}&\footnotesize{(82.8-99.8)}&\footnotesize{(63.5-89.0)}\\
\hline
CC& & & & & & & & \\
\hspace{0.5cm}\bf{\footnotesize{LR}}&\bf{\footnotesize{0.8$\pm$6.0}}&\bf{\footnotesize{0.92$\pm$0.05}}&\bf{\footnotesize{1.7$\pm$12.4}}&\bf{\footnotesize{100.0$\pm$0.0}}&\bf{\footnotesize{0.7$\pm$8.4}}&\bf{\footnotesize{0.6$\pm$5.2}}\\
&\bf{\footnotesize{(0.4-1.2)}}&\bf{\footnotesize{(0.92-0.92)}}&\bf{\footnotesize{(0.8-2.5)}}&\bf{\footnotesize{(100.0-100.0)}}&\bf{\footnotesize{(0.1-1.3)}}&\bf{\footnotesize{(0.2-1.0)}}\\
\hspace{0.5cm}\footnotesize{SVM}&\footnotesize{0.6$\pm$3.9}&\footnotesize{0.76$\pm$0.32}&\footnotesize{0.7$\pm$7.4}&\footnotesize{100.0$\pm$0.0}&\footnotesize{0.9$\pm$9.3}&\footnotesize{0.3$\pm$0.0}\\
&\footnotesize{(0.4-0.9)}&\footnotesize{(0.74-0.79)}&\footnotesize{(0.2-1.3)}&\footnotesize{(100.0-100.0)}&\footnotesize{(0.1-1.7)}&\footnotesize{(0.3-0.3)}\\
\hspace{0.5cm}\footnotesize{NB}&\footnotesize{0.6$\pm$3.2}&\footnotesize{0.92$\pm$0.05}&\footnotesize{1.1$\pm$9.4}&\footnotesize{100.0$\pm$0.0}&\footnotesize{1.0$\pm$9.9}&\footnotesize{0.3$\pm$0.0}\\
&\footnotesize{(0.4-0.9)}&\footnotesize{(0.91-0.92)}&\footnotesize{(0.4-1.8)}&\footnotesize{(100.0-100.0)}&\footnotesize{(0.3-1.7)}&\footnotesize{(0.3-0.3)}\\
\hspace{0.5cm}\footnotesize{RF}&\footnotesize{0.3$\pm$0.0}&\footnotesize{0.48$\pm$0.06}&\footnotesize{0.0$\pm$0.0}&\footnotesize{100.0$\pm$0.0}&\footnotesize{0.0$\pm$0.0}&\footnotesize{0.3$\pm$0.0}\\
&\footnotesize{(0.3-0.3)}&\footnotesize{(0.47-0.50)}&\footnotesize{(0.0-0.0)}&\footnotesize{(100.0-100.0)}&\footnotesize{(0.0-0.0)}&\footnotesize{(0.3-0.3)}\\
\hline
MI& & & & & & & & \\
\hspace{0.5cm}\footnotesize{LR}&\footnotesize{30.3$\pm$40.8}&\footnotesize{0.97$\pm$0.06}&\footnotesize{30.1$\pm$41.0}&\footnotesize{100.0$\pm$0.0}&\footnotesize{38.9$\pm$48.8}&\footnotesize{20.8$\pm$40.2}\\
&\footnotesize{(27.4-33.2)}&\footnotesize{(0.96-0.97)}&\footnotesize{(27.2-33.0)}&\footnotesize{(100.0-100.0)}&\footnotesize{(35.4-42.4)}&\footnotesize{(18.0-23.7)}\\
\hspace{0.5cm}\footnotesize{SVM}&\footnotesize{8.9$\pm$24.9}&\footnotesize{0.54$\pm$0.44}&\footnotesize{8.6$\pm$24.9}&\footnotesize{100.0$\pm$0.0}&\footnotesize{12.6$\pm$33.2}&\footnotesize{5.3$\pm$21.7}\\
&\footnotesize{(7.2-10.7)}&\footnotesize{(0.50-0.57)}&\footnotesize{(6.9-10.4)}&\footnotesize{(100.0-100.0)}&\footnotesize{(10.2-15.0)}&\footnotesize{(3.8-6.9)}\\
\hspace{0.5cm}\bf{\footnotesize{NB}}&\bf{\footnotesize{30.9$\pm$41.0}}&\bf{\footnotesize{0.97$\pm$0.07}}&\bf{\footnotesize{30.9$\pm$41.3}}&\bf{\footnotesize{100.0$\pm$0.0}}&\bf{\footnotesize{39.8$\pm$49.0}}&\bf{\footnotesize{21.0$\pm$40.4}}\\
&\bf{\footnotesize{(28.0-33.8)}}&\bf{\footnotesize{(0.96-0.97)}}&\bf{\footnotesize{(28.0-33.8)}}&\bf{\footnotesize{(100.0-100.0)}}&\bf{\footnotesize{(36.3-43.4)}}&\bf{\footnotesize{(18.1-23.9)}}\\
\hspace{0.5cm}\footnotesize{RF}&\footnotesize{27.9$\pm$40.2}&\footnotesize{0.75$\pm$0.23}&\footnotesize{27.6$\pm$40.4}&\footnotesize{100.0$\pm$0.0}&\footnotesize{36.7$\pm$48.5}&\footnotesize{20.6$\pm$40.3}\\
&\footnotesize{(18.9-36.9)}&\footnotesize{(0.70-0.80)}&\footnotesize{(18.6-36.7)}&\footnotesize{(100.0-100.0)}&\footnotesize{(25.8-47.6)}&\footnotesize{(11.5-29.6)}\\
\hline
\end{tabular*}
\caption{Classification scores of the various classifiers on the ``Real testing dataset''. Quantitative scores were derived via evaluation of RegQCNET, correlation coefficient (CC), and mutual information (MI) (after 10-fold cross-validation) by various machine-learning algorithms (LR: logistic regression, SVM: support machine vector, NB: na\"ive bayes, RF: random forest). Quantitative indicators are shown with standard deviations and 95$\%$ confidence intervals in parentheses. Best performance are reported in bold font for each indicator.}
\label{table:classification}
\end{table}

\section{Discussion}

The proposed method aims at quantifying the amplitude of the spatial affine mismatch between a brain MRI and a template. To this end, we used RMSE as criterion to evaluate affine registration quality. Our experimental results demonstrate that the proposed RegQCNET outperforms traditional intensity-based criteria. Moreover, using automatic threshold, our approach delivers reproducible results and minimizes operator dependency.

It can be observed that good scores were obtained thanks to the use of i) a well-balanced training population covering the entire lifespan and ii) a well-balanced distribution of the simulated transformations (\emph{i.e.}, uniform distribution of the resulting RMSEs). It is interesting to highlight that comparable precision and accuracy were obtained on simulated data (as described in section \ref{sssec:test1}) using a subsampling factor 2 on the images (instead of the subsampling factor 4 used in the presented results).

One can distinguish two potential contributions on the apparent image-to-template mismatch: (i) the effective registration error that we aim to quantify and (ii) the anatomical variability between the actual image and the template (in particular, the size of the brain varies a lot along the lifespan). 
Any unobserved spatial transformations/brain shapes/image artifacts during the training step may disturb in turn the proposed quantitative CNN-based QC. This phenomenon can be observed in Fig. \ref{fig:RMSE}a where an insufficiently populated training dataset was employed (100 images). In turn, a dramatic impact arise on both precision and accuracy. Note that using correlation or mutual information as registration QC, any intensity variation between a given brain MRI and the template is attributed to an image mismatch. The tested machine learning classifiers thus provided very poor results in terms of accuracy, sensitivity, PPV and NPV.

Another limitation arise when an intensity perturbation occurs between training and testing. While a spatially homogeneous bias did not impact the performance (Fig. \ref{fig:intensity_bias}c), an accurate correction of spatially heterogeneous intensity bias is a necessary prerequisite to obtain good performance when using RegQCNET (Fig. \ref{fig:intensity_bias}e) \cite{preprocessing3} \cite{MRI_normalisation}. It has to be noted that our framework was robust to various degrees of masking due to defacing (\emph{e.g.}, NDAR dataset).

As one can expect, the range of spatial transformations during training has to be carefully determined. That brings us to an inherent limit of the proposed technique. Indeed, a RMSE extrapolation outside training limits is intrinsically not possible using our CNN-based approach and thus a large training range is mandatory. This limitation could be limited by training RegQCNET on a larger range of RMSE.

Using the proposed CNN-based QC, a few tenth of a seconds ($700$ ms and $200$ ms without and with the use of GPU acceleration, respectively) is needed to provide a quantitative prediction of the image-to-template alignment accuracy. This perfectly meets our computational requirements related to the inclusion of this QC step in massive processing.

While our experimental results demonstrate that a compact network is an efficient solution to estimate the quality of affine registration between a T1-MRI and a template, the direct translation of our approach to radiation therapies is not straightforward. 
In the context of radiation therapies, several performance indicators and registration QC solutions have been proposed (see \cite{RegQC_RT_3} and \cite{RegQC_RT_1}). Concerning abdominal organs, it must be underlined that deformations are non-rigid and thus extension of our framework to non-rigid registration would be required. For instance, recent works conducted in the abdomen used biomechanical criteria to assess image registration accuracy \cite{QA_Zachiu_2018} \cite{QA_Zachiu_2020}. Such approaches involve the estimation of mechanical stress, which would occur within the observed tissues. The calculated stress can then be compared to plausible physiological limits. We believe that a combination of these two complementary approaches (\emph{i.e.}, the network and the biomechanical strategies) should be investigated in future studies.

\section{Conclusion}

This study demonstrates that quantitative estimation of registration mismatch between a brain image and a template can be achieved using 3D CNNs. However, to ensure the quality of the estimation, the training dataset have to be carefully designed. To this end, in this study we used: i) a gender and age well-balanced lifespan dataset covering the entire lifespan, ii) an uniformly distributed amplitudes of random spatial transformations to cover registration error from 0 to 100 millimeters, and iii) a sufficient amplitude range of simulated spatial transformations. The proposed tool can be used as quality control for automated image registration of T1-weigthed brain onto a reference template.

Future studies will include the extension of the proposed RegQCNET to complex elastics image deformations, the estimation of 3D RMSE maps, the impact of incomplete, noisy and corrupted brains, as well as the extension of the method to cross-contrast and multi-modal images.

\section*{Acknowledgment}

Experiments presented in this paper were carried out using the PlaFRIM experimental testbed, supported by Inria, CNRS (LABRI and IMB), Universit\'e de Bordeaux, Bordeaux INP and Conseil R\'egional d'Aquitaine (see https://www.plafrim.fr/). This work benefited from the support of the project DeepvolBrain of the French National Research Agency (ANR-18-CE45-0013). This study was achieved within the context of the Laboratory of Excellence TRAIL ANR-10-LABX-57 for the BigDataBrain project. Moreover, we thank the Investments for the future Program IdEx Bordeaux (ANR-10-IDEX-03-02, HL-MRI Project), Cluster of excellence CPU and the CNRS/INSERM for the DeepMultiBrain project. This study has been also supported by the DPI2017-87743-R grant from the Spanish Ministerio de Economia, Industria Competitividad. The authors gratefully acknowledge the support of NVIDIA Corporation with their donation of a TITAN X GPU used in this research.

\section*{References}

\bibliographystyle{dcu}
\bibliography{2020_RegQCNET_revision2_v1_clean_copie}

\end{document}